\documentclass[a4paper, oneside, reqno, 11pt]{amsart}

\usepackage{lipsum}
 \let\counterwithin\relax \usepackage{graphicx,amsfonts,amssymb,amsmath,amsthm,url,amscd}  \usepackage{color} \usepackage{chngcntr}
\usepackage[dvipsnames]{xcolor} \usepackage[normalem]{ulem} \usepackage{comment}
\usepackage{graphicx, amsmath, amssymb, amsfonts, amsthm, tikz, amscd,mathtools}
\usetikzlibrary{matrix,arrows,decorations.pathmorphing}
\usepackage[a4paper, margin=2.7cm]{geometry}
\usepackage{verbatim}
\usepackage{calc}
\usepackage{dsfont}
\usepackage{enumerate}

\usepackage[labelformat=simple]{subcaption}

\makeatletter
\renewcommand\p@subfigure{}
\makeatother

\usepackage{pdflscape}

\usepackage[linesnumbered, ruled, vlined, algosection]{algorithm2e}
\SetKwInput{KwComplexity}{Complexity}
\SetKw{Continue}{continue}
\SetKw{False}{false}
\SetKw{True}{true}
\SetKw{And}{and}
\SetKw{Or}{or}
\SetKw{Delete}{delete}
\SetKw{Not}{not}
\SetKw{Then}{then}

\counterwithin{figure}{section}

\usepackage{placeins}

\let\Oldsection\section
\renewcommand{\section}{\FloatBarrier\Oldsection}

\let\Oldsubsection\subsection
\renewcommand{\subsection}{\FloatBarrier\Oldsubsection}

\let\Oldsubsubsection\subsubsection
\renewcommand{\subsubsection}{\FloatBarrier\Oldsubsubsection}

\usepackage{fewerfloatpages}

\usepackage{tabularray}

\usepackage[font = footnotesize]{caption}

\UseTblrLibrary{siunitx}
\UseTblrLibrary{counter}
\UseTblrLibrary{functional}

\DefTblrTemplate{caption-tag}{caption-ams-style}{\textsc{Table~\thetable .}}
\DefTblrTemplate{caption-sep}{caption-ams-style}{\enskip}
\DefTblrTemplate{caption-text}{caption-ams-style}{\InsertTblrText{caption}}
\DefTblrTemplate{conthead-text}{caption-ams-style}{\mbox{(Continued~on~next~page)}}

\ExplSyntaxOn
\DefTblrTemplate{caption}{caption-ams-style}{\vspace{0.7em}
	\hbox_set:Nn \l__tblr_caption_box
	{
		\UseTblrTemplate { caption-tag } { default }
		\UseTblrTemplate { caption-sep } { default }
		\UseTblrTemplate { caption-text } { default }
	}
	\dim_compare:nNnTF { \box_wd:N \l__tblr_caption_box } > { 0.85\linewidth }
	{
		\noindent
		\makebox[\linewidth][c]{
			\parbox{0.85\textwidth}{
				\hbox_unpack:N \l__tblr_caption_box
			}
		}
		\par
	}
	{
		\centering
		\makebox [\linewidth] [c] { \box_use:N \l__tblr_caption_box }
		\par
	}
}
\DefTblrTemplate{capcont}{caption-ams-style}{\vspace{0.7em}
	\hbox_set:Nn \l__tblr_caption_box
	{
		\UseTblrTemplate { caption-tag } { default }
		\UseTblrTemplate { caption-sep } { default }
		\UseTblrTemplate { caption-text } { default }
		\UseTblrTemplate { conthead-pre } { default }
		\UseTblrTemplate { conthead-text } { default }
	}
	\dim_compare:nNnTF { \box_wd:N \l__tblr_caption_box } > { 0.85\linewidth }
	{
		\noindent
		\makebox[\linewidth][c]{
			\parbox{0.85\textwidth}{
				\hbox_unpack:N \l__tblr_caption_box
			}
		}
		\par
	}
	{
		\centering
		\makebox [\linewidth] [c] { \box_use:N \l__tblr_caption_box }
		\par
	}
}
\ExplSyntaxOff

\DefTblrTemplate{firsthead,middlehead,lasthead}{caption-bottom-ams-style}{
}
\DefTblrTemplate{firstfoot}{caption-bottom-ams-style}{
	\UseTblrTemplate{capcont}{caption-ams-style}
}
\DefTblrTemplate{middlefoot}{caption-bottom-ams-style}{
	\UseTblrTemplate{capcont}{caption-ams-style}
}
\ExplSyntaxOn
\DefTblrTemplate{lastfoot}{caption-bottom-ams-style}{
	\UseTblrTemplate{note}{default}
	\UseTblrTemplate{remark}{default}
		\UseTblrTemplate{caption}{caption-ams-style}
}
\ExplSyntaxOff

\NewTblrTheme{ams-theme}{
	\SetTblrTemplate{caption-tag}{caption-ams-style}
	\SetTblrTemplate{caption-sep}{caption-ams-style}
	\SetTblrTemplate{caption-text}{caption-ams-style}
	
	\SetTblrTemplate{conthead-text}{caption-ams-style}
	\SetTblrTemplate{caption}{caption-ams-style}
	
	\SetTblrTemplate{firsthead}{caption-bottom-ams-style}
	\SetTblrTemplate{middlehead}{caption-bottom-ams-style}
	\SetTblrTemplate{lasthead}{caption-bottom-ams-style}
	\SetTblrTemplate{firstfoot}{caption-bottom-ams-style}
	\SetTblrTemplate{middlefoot}{caption-bottom-ams-style}
	\SetTblrTemplate{lastfoot}{caption-bottom-ams-style}
}

\counterwithin{table}{section}

\usepackage{etoolbox} \patchcmd{\section}{\scshape}{\bfseries}{}{} \makeatletter \renewcommand{\@secnumfont}{\bfseries} \makeatother
\newrobustcmd\TableBold{\DeclareFontSeriesDefault[rm]{bf}{b}\bfseries}

\usepackage{graphicx, amsmath, amssymb, amsfonts, amsthm, stmaryrd, tikz, amscd} \usepackage[all]{xy} \usetikzlibrary{matrix,arrows,decorations.pathmorphing}

\usepackage[hyperfootnotes=false, colorlinks, citecolor= RoyalBlue, urlcolor=blue, linkcolor=blue ]{hyperref}

\makeatletter
\let\orgdescriptionlabel\descriptionlabel
\renewcommand*{\descriptionlabel}[1]{%
  \let\orglabel\label
  \let\label\@gobble
  \phantomsection
  \protected@edef\@currentlabel{#1}%
  \let\label\orglabel
  \orgdescriptionlabel{#1}%
}
\def\paragraph{
	\@startsection{paragraph}{4}
	\z@{.5\linespacing\@plus.7\linespacing}{-.5em}%
	{\normalfont\itshape}}
\makeatother

\theoremstyle{plain} 
\newtheorem{theorem} {Theorem}

\theoremstyle{definition}

\theoremstyle{plain} 

\theoremstyle{remark}

\newtheorem{definition}[theorem]{Definition}
\newtheorem*{definition*} {Definition}
\newtheorem*{theorem*} {Theorem}

 \newtheorem*{example*} {Example}

\newtheorem{remark}[theorem] {Remark} \newtheorem*{remark*} {Remark}       

\newtheoremstyle{itplain} 
{6pt} 
{5pt\topsep} 
{\itshape} 
{} 
{\itshape} 
{.}  
{5pt plus 1pt minus 1pt} 
{} 

\theoremstyle{itplain} 

\newtheorem*{lemma*}{Lemma}

\newtheorem*{corollary*} {Corollary}

\theoremstyle{remark} 

\newtheorem*{lemmatest*}{Lemma}

\usepackage{thmtools, thm-restate}

\usepackage{etoolbox} \patchcmd{\section}{\scshape}{\bfseries}{}{} \makeatletter \renewcommand{\@secnumfont}{\bfseries} \makeatother

 \renewcommand{\leq}{\leqslant} 

\newlength{\faktorheight}

\makeatletter
\providecommand{\leftsquigarrow}{%
	\mathrel{\mathpalette\reflect@squig\relax}%
}
\newcommand{\reflect@squig}[2]{%
	\reflectbox{$\m@th#1\rightsquigarrow$}%
}
\makeatother

\numberwithin{theorem}{section}
\numberwithin{equation}{section}                                 \def\eps{\varepsilon}

\DeclareMathOperator*{\argmin}{argmin}

\newcommand{\sm}{\left(\begin{smallmatrix}} \newcommand{\esm}{\end{smallmatrix}\right)} \newcommand{\bpm}{\begin{pmatrix}} \newcommand{\ebpm}{\end{pmatrix}}

\newcommand{\vect}[1]{{\boldsymbol{#1}}}

\newcommand{\lquotient}[2]{ \mathchoice {
    \text{\lower1ex\hbox{$#1$}\Big \backslash \raise01ex\hbox{$#2$}}%
  } {
    #1\,\backslash\,#2 } {
    #1\,\backslash\,#2 } {
    #1\,\backslash\,#2 } }

\newcommand{\rquotient}[2]{ \mathchoice {
    \text{\raise01ex\hbox{$#1$}\Big/\lower1ex\hbox{$#2$}}%
  } {
    #1\,/\,#2 } {
    #1\,/\,#2 } {
    #1\,/\,#2 } }
		
\newcommand{\lrquotient}[3]{ \mathchoice {
    \text{\lower1ex\hbox{$#1$}\Big \backslash \raise01ex\hbox{$#2$}\Big/\lower1ex\hbox{$#3$}}%
  } {
    #1\,\backslash\,#2\,/\,#3 } {
    #1\,\backslash\,#2\,/\,#3 } {
    #1\,\backslash\,#2\,/\,#3 } }

\author{Dimosthenis Pasadakis\textsuperscript{\textdagger}}
\address[Dimosthenis Pasadakis]{Institute of Computing, Universit\`{a} della Svizzera italiana, Via la Santa 1, 6962 Lugano, Switzerland} \email{dimosthenis.pasadakis@usi.ch}

\author{Raphael S. Steiner\textsuperscript{\textdagger}}
\address[Raphael S. Steiner]{Huawei Research Center Zurich, Computing Systems Lab, Thurgauerstrasse 80, 8050 Z{\"u}rich, Switzerland} \email{raphael.steiner@huawei.com}

\author{P\'al Andr\'as Papp}
\address[P\'al Andr\'as Papp]{Huawei Research Center Zurich, Computing Systems Lab, Thurgauerstrasse 80, 8050 Z{\"u}rich, Switzerland} \email{pal.andras.papp@huawei.com}

\author{Toni B\"ohnlein}
\address[Toni B\"ohnlein]{Huawei Research Center Zurich, Computing Systems Lab, Thurgauerstrasse 80, 8050 Z{\"u}rich, Switzerland} \email{toni.boehnlein@huawei.com}

\author{Albert-Jan N. Yzelman}
\address[Albert-Jan N. Yzelman]{Huawei Research Center Zurich, Computing Systems Lab, Thurgauerstrasse 80, 8050 Z{\"u}rich, Switzerland} \email{albertjan.yzelman@huawei.com}

\date{\today}
\title[Symmetry-breaking symmetry in directed spectral partitioning]{Symmetry-breaking symmetry \\ in directed spectral partitioning}


\hypersetup{ pdfkeywords={}, pdfsubject={}, pdfcreator={TeXstudio 2.12.22}}

\begin{document}

\begin{abstract} We break the symmetry in classical spectral bi-partitioning in order to incentivise the alignment of directed cut edges. We use this to generate acyclic bi-partitions and furthermore topological orders of directed acyclic graphs with superb locality. The new approach outperforms the state-of-the-art Gorder algorithm by up to $17\times$ on total reuse distance and minimum linear arrangement.
\end{abstract}

\maketitle
\renewcommand*{\thefootnote}{\fnsymbol{footnote}}
\footnotetext[2]{Joint first authors; listed in alphabetical order.}
\renewcommand*{\thefootnote}{\arabic{footnote}}

\setcounter{tocdepth}{1} \tableofcontents


\section{Introduction}\label{sec:introduction}

Graph partitioning plays a central role in a variety of applications, for example classification, recommendation systems, parallel processing, very-large-scale integration (VLSI), and routing to name a few. The basic premise is often the same: identifying tightly knit groups of roughly the same size with few interconnects. To this end, a plethora of methods and algorithms have been developed ranging from geometric, spectral, local-search, multi-level, evolutionary methods, and combinations of the aforementioned. We refer to the surveys \cite{fjallstrom1998algorithms, schloegel2000graph, monien2007approximation, kim2011genetic, bichot2013graph, bulucc2016recent, ccatalyurek2023more}.

A specialisation of the graph partitioning problem is acyclic partitioning, where one is given a directed acyclic graph and the partition itself is required to form a directed acyclic graph\footnote{Specifically the quotient graph with self-loops removed.}. This stricter problem is required, for example, when the directed acyclic graph is modelling a computation with dependencies. As such, it has been used for pipeline computations, efficient uses of the memory hierarchy, and, in general, as a component of offline scheduling heuristics \cite{agrawal2012cache, fauzia2013beyond, elango2015characterizing, ozkaya2019acyclic, papp24a}.

Between undirected and acyclic graph partitioning, there also lies a more relaxed partitioning problem: given a directed graph, one seeks to find a partition such that between any two parts the cut edges mostly point in one direction. This type of partitioning can be used as initialisers for acyclic partitions and for flow-based clustering \cite{hayashi2022skew}, such as hierarchy classifications \cite{van2019spectral}, interbank debts~\cite{Acemoglu15}, and population migration patterns~\cite{zhang25}. 

In this paper, we present a modification to the classic spectral partitioning method of Fiedler \cite{fiedler1973algebraic, fiedler1989laplacian} which yields an aforementioned nearly acyclic bi-partition. In the case of a directed acyclic graph, we furthermore augment the nearly acyclic bi-partition to an acyclic bi-partition. To the best of our knowledge, this marks the first time spectral methods have been used to address the problem of acyclic partitioning.

As the final and central application of our method in this paper, we turn to the graph-layout problem, also known as the linear-arrangement problem, of directed acyclic graphs. The latter asks for a topological order with `good locality'. Specifications of locality take on a variety of shapes. The most well-known metrics to evaluate the locality of an ordering include the bandwidth, cut width, minimum linear arrangement (sum of edge lengths), and reuse distance. The associated problems have been well-studied, both from a theoretical and a practical point of view, as they are widely applicable. Applications range from VLSI design, single-core and multi-core scheduling, information retrieval, cache utilisation, network reliability, to speeding up graph and sparse matrix algorithms \cite{chung1988labelings, diaz2002survey, csc12, petit2013addenda}. 

\begin{figure}[!htpb]
	\centering
	
	\subcaptionbox{\label{fig:barth-spectral} Spectral-dir-top}%
	{\includegraphics[width=0.4\textwidth]{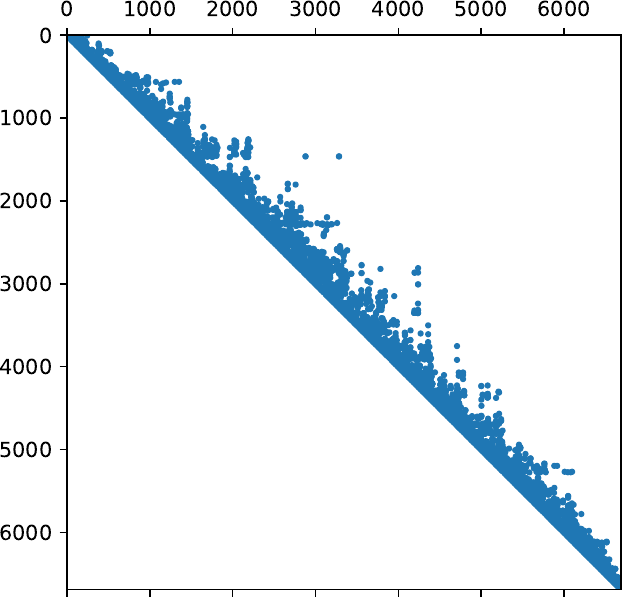}}
	\hspace*{0.05\textwidth}	
	\subcaptionbox{\label{fig:barth-gorder} Gorder}%
	{\includegraphics[width=0.4\textwidth]{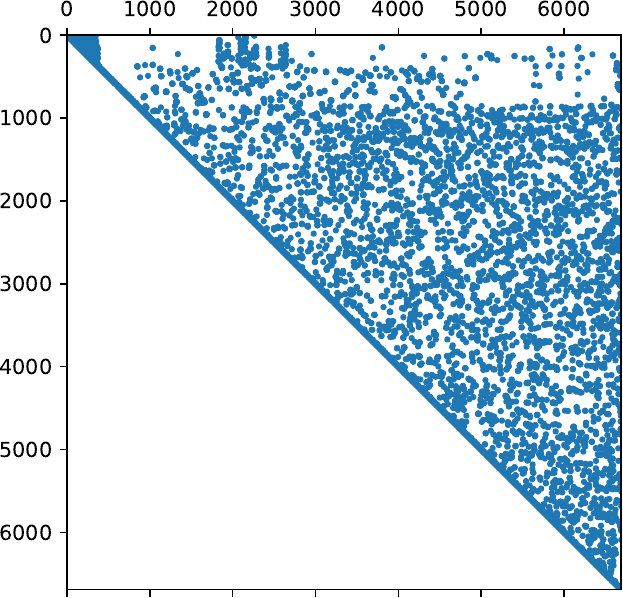}}
	\caption{\label{fig:intro-matrix-comparison} Topological orderings generated by \ref{fig:barth-spectral} our direction-incentivised spectral topological-order method (Spectral-dir-top) and \ref{fig:barth-gorder} an acyclic adaptation of Gorder (Gorder). The directed acyclic graph corresponds to the lower triangular part of the matrix `barth'.}
\end{figure}

In due course, we will show that our method excels at generating topological orders for data and temporal locality. Indeed, we report significant improvements over an acyclic adaptation of the state-of-the-art method Gorder \cite{wei2016speedup}: up to $28\times$ smaller minimal linear arrangement, up to $17\times$ smaller total reuse distance, $11\times$ smaller cut width, and $6 \times $ smaller bandwidth and maximum reuse distance. Figure~\ref{fig:intro-matrix-comparison} shows an example with $10\times$ smaller minimal linear arrangement and total reuse distance.

\subsection{Nearly acyclic bi-partition}

There are several spectral methods in the literature that deal with directed graphs. Broadly speaking, they can be categorised into two camps: flow-based and density-based \cite{hayashi2022skew}. Examples for flow-based techniques include clustering on an embedding based on spectral information of the (dampened) page-rank transition matrix \cite{van2019spectral} or Hermitian variants \cite{guo2017hermitian, mohar2020new, laenen2020higher}, bibliographic coupling matrix \cite{kessler1963bibliographic}, co-citation matrix \cite{small1973co}, and linear combinations thereof \cite{satuluri2011symmetrizations}. Their main feature is that they group vertices together if they have a lot of in- and/or out-neighbours in common, but disregard grouping vertices together which are tightly knit. This makes them suitable for detecting hierarchies and flows, but unsuitable for our purposes. On the other hand, density-based techniques such as using the directed Laplacian \cite{zhou2005learning, huang2006web, gleich2006hierarchical} group tightly knit vertices, but generally fail to have most cut edges between groups aligned.

In \S\ref{sec:partition} of this paper, we address this gap. We achieve this by taking density-based spectral partitioning methods, such as the undirected Laplacian \cite{fiedler1973algebraic, fiedler1989laplacian}, $p$-Laplacian \cite{buhler2009spectral, simpson2018balanced}, or the directed Laplacian \cite{chung2005laplacians}, and modify them as to incentivise the alignment of cut edges. The general idea is as follows. Vaguely speaking, the aforementioned methods assign a vertex the value $1$ if they belong to the first cluster and $-1$ if they belong to the second cluster. The value of a directed edge is given as the (signed) difference of the source and target of the edge. Thus, the value is $0$ for a non-cut edge, $2$ for a cut edge from the first cluster to the second, and $-2$ for a cut edge pointing in the reverse direction. Therefore, the product of the values of two cut edges tells us now whether they are aligned or not. Hence, summing up all these products over all pairs of edges gives us a term which penalises non-aligned cut edges. As this penalty term constitutes a quadratic form, we are able to incorporate it into spectral partitioning methods.

\subsection{Acyclic bi-partition}

Like balanced undirected bi-partitioning, balanced acyclic bi-partitioning is NP-hard \cite{garey1974some, moreiraAcyclicPartition}. 
Yet, the latter is provably harder in the following sense. For $\eps$-balanced\footnote{All parts of the partition are at most a factor $(1+\eps)$ larger than their mean, cf.\@ Definition \ref{def:balanced-partition}.} undirected bi-partitioning, there exists a polynomial-time $O(\log(|V|))$-approximation algorithm by R\"acke \cite{racke2008optimal}, whereas we show that no such algorithm exists for $\eps$-balanced acyclic bi-partitioning under the exponential-time hypothesis \cite{impagliazzo2001complexity}.

\begin{restatable}{theorem}{thmInapprox}
	\label{thm:acyclic-bi-partition-inapprox}%
	Assuming the exponential-time hypothesis, there is a $\delta > 0$ such that for every $0 \le \eps < 1$ the $\eps$-balanced acyclic bi-partition problem does not have a polynomial-time $n^{1/\log(\log(n))^{\delta}}$-factor approximation algorithm, where $n = |V|$ is the size of the input graph.
%
\end{restatable}

Despite the NP-hardness result, there are a variety of fast heuristics and algorithms in the literature ranging from topological-order-based \cite{kernighan1971optimal}, to acyclic adaptations of the Fiduccia--Mattheyses method \cite{cong1994acyclic, popp2021multilevel}, evolutionary algorithms \cite{sanders2017distributed, moreira2018evolutionary}, and multi-level algorithms \cite{herrmann2017acyclic, herrmann2019multilevel, popp2021multilevel}. We refer to the survey \cite{ccatalyurek2023more} and the references therein for further information.

In \S\ref{sec:acyc-bi-partition} of this paper, we present an acyclic bi-partitioner based on our nearly acyclic spectral bi-partitioner. To fix the acyclicity, we generate a topological order of the directed acyclic graph, which aims to preserve the bi-partition, and subsequently bisect said topological order. To the best of our knowledge, this marks the first acyclic bi-partitioner based on spectral methods.


\subsection{Directed acyclic graph layout}


A multitude of different techniques and heuristics have been developed to address graph-layout or linear-arrangement problems. Methods range from (approximate) minimal degree orderings \cite{amestoy96,amestoy04}, (reverse) Cuthill--Mckee \cite{cuthill1969reducing,liu76}, highest recent relations heuristics such as Gorder \cite{wei2016speedup}, and recursively partitioning-based \cite{schamberger2004locality}. For a broader overview of the literature, we refer to \cite{chung1988labelings, diaz2002survey, csc12, petit2013addenda}.

In \S\ref{sec:spec-top-order} of this paper, we present a heuristic based on our spectral acyclic bi-partition algorithm for the graph-layout problem for directed acyclic graphs with the additional requirement that the linear arrangement must be a topological order. Our heuristic works similar to the one given by Schamberger--Wierum \cite{schamberger2004locality} for undirected graphs in the sense that it recursively (acyclicly bi-)partitions the graph into smaller parts which are then laid out in a (topological) order. We note that similar schemes have been employed to address numerous other problems, cf.\@ \cite{yzelman09,grigori2010hypergraph,yzelman11a,abubaker18,mondriaan}. Here, we make a subtle but notable change to these schemes. Namely, when we acyclicly bi-partition a subgraph, we take the whole graph into account and not just the subgraph. This allows us to not only minimise the edge lengths of the edges within the subgraph, but also the ones that have already been cut by previous rounds of acyclic bi-partitioning. This leads to an overall better graph layout.

\subsection{Overview} In \S\ref{sec:partition}, we discuss the spectral bi-partitioning method with cut-edge direction incentive; in \S\ref{sec:acyc-bi-partition}, we extend the method to produce acyclic bi-partitions; and in \S\ref{sec:spec-top-order}, we discuss how this can be used to generate a topological order with good locality properties. The evaluation of every method may be found in its respective section. In Appendix \ref{sec:directed-mincut-inapprox}, we give the proof of the inapproximability result, Theorem \ref{thm:acyclic-bi-partition-inapprox}. 

\subsection{Acknowledgements}
We thank Davide B.\@ Bartolini and Mehdi Alipour for stimulating conversations on this and neighbouring topics. We are furthermore thankful to Davide B.\@ Bartolini for comments on an earlier draft.

Dimosthenis Pasadakis was financially supported by the joint DFG-470857344 and SNSF-204817 project, and by the Huawei Zurich Research Center.

\section{Nearly acyclic spectral bi-partitioning}\label{sec:partition}

\subsection{Algorithm} \label{sec:bi-part-alg}

Classical spectral bi-partitioning is a continuous relaxation of the discrete min-cut problem on an undirected graph $G=(V,E)$. The problem may be stated as
\begin{equation} \label{eq:classic-spec-partition}
\argmin_{\substack{ \vect{x} \in \mathbb{R}^V \\ \|\vect{x}\|_2 = 1 \\ \vect{x} \perp \vect{1}  }} \sum_{(u,v) \in E} ( x_u - x_v )^2
\end{equation}
and its solution is given by the Fiedler eigenvector \cite{fiedler1973algebraic, fiedler1989laplacian}. We refer to \cite{chung1997spectral} and references therein for an extensive overview of the literature.

In this paper, we consider a \emph{directed} graph $G$ and would like the cut edges to mostly align in the same direction. We achieve this by introducing a penalty to \eqref{eq:classic-spec-partition} when the cut edges are not aligned.

\subsubsection{Symmetry-breaking quadratic form} \label{sec:quad-form-partition}

Due to the symmetry of the quadratic form, we are unable to enforce that a cut edge $x_u - x_v\not \approx 0$, $(u,v)\in E$, is positive. However, by considering pairs of edges, we are able to construct a quadratic form which is large if and only if all cut edges point in the same direction; that is, they have the same sign:
\begin{equation} \label{eq:penalty}
\sum_{\substack{ (u_1,v_1) \in E \\ (u_2,v_2) \in E }} (x_{u_1} - x_{v_1})(x_{u_2} - x_{v_2}).
\end{equation}

We note that by Cauchy--Schwarz, we have
\begin{equation} \label{eq:upper-bnd-penalty}
|E| \cdot \sum_{(u,v) \in E} ( x_u - x_v )^2 \ge \left( \sum_{(u,v) \in E}  (x_u - x_v) \right)^2 = \sum_{\substack{ (u_1,v_1) \in E \\ (u_2,v_2) \in E }} (x_{u_1} - x_{v_1})(x_{u_2} - x_{v_2}).
\end{equation}
Hence,
\begin{equation}
\label{eq:symmetry-breaking-quad-form}
\sum_{(u,v) \in E} ( x_u - x_v )^2 - c\left( \sum_{(u,v) \in E}  (x_u - x_v) \right)^2
\end{equation}
is a symmetric positive semi-definite quadratic form for $0 \le c \le \frac{1}{|E|}$, which we can use for spectral bi-partitioning. More precisely, we can squeeze the quadratic form \eqref{eq:symmetry-breaking-quad-form} between multiples of the quadratic form \eqref{eq:classic-spec-partition} used in classic spectral partitioning.

\begin{equation} \label{eq:squeeze-penalised-quadratic-form}
	\sum_{(u,v) \in E} ( x_u - x_v )^2 \ge \sum_{(u,v) \in E} ( x_u - x_v )^2 - c\left( \sum_{(u,v) \in E}  (x_u - x_v) \right)^2 \ge \left(1 - c|E|\right) \sum_{(u,v) \in E} ( x_u - x_v )^2.
\end{equation}

For $0 \le c < \frac{1}{|E|}$, it follows that \eqref{eq:symmetry-breaking-quad-form} is zero if and only if the Laplacian \eqref{eq:classic-spec-partition} is zero. Hence, like the Laplacian, the kernel of quadratic form \eqref{eq:symmetry-breaking-quad-form} identifies the weakly connected components of the graph $G$. It furthermore follows that theoretical results on classic spectral partitioning on cut quality continue to hold with an additional multiplicative loss of at most $(1-c|E|)^{-1}$, e.g., (undirected) Cheeger's inequality \cite{alon1985lambda1, alon1986eigenvalues, sinclair1989approximate}.

The final optimisation problem reads:
\begin{equation} \label{eq:symmetry-breaking-optimisation-problem}
\argmin_{\substack{ \vect{x} \in \mathbb{R}^V \\ \|\vect{x}\|_2 = 1 \\ \vect{x} \perp \vect{1}  }}
\sum_{(u,v) \in E} ( x_u - x_v )^2 - c\left( \sum_{(u,v) \in E}  (x_u - x_v) \right)^2,
\end{equation}
for some $0 \le c \le \frac{1}{|E|}$.

\begin{remark}
	The above ideas and methods are easily generalised to (non-negatively) edge-weighted graphs. We leave the details to the interested reader.
\end{remark}

\begin{remark}
	We have
	\begin{equation}
		\sum_{(u,v) \in E}  (x_u - x_v) = \sum_{v \in V} \bigl(d^+(v) - d^-(v) \bigr) x_v,
	\end{equation}
	where $d^+(v)$ denotes the out-degree and $d^-(v)$ the in-degree of a vertex $v \in V$. Hence, the alignment incentive can also be seen as the following heuristic: put vertices with larger difference of out-degree and in-degree in one part and the ones with smaller difference in the other.
\end{remark}

\begin{remark}[$p$-Laplacian] \label{rem:p-Laplacian-partition}

It has been demonstrated that using the $p$-Laplacian for $p$ close to but strictly larger than $1$ leads to better cut quality \cite{buhler2009spectral, simpson2018balanced}.
One may generalise our penalty term \eqref{eq:penalty} also to this setting. The optimisation problem reads now as follows:
\begin{equation} \label{eq:p-spec-partition}
\argmin_{\substack{ \vect{x} \in \mathbb{R}^V \\ \|\vect{x}\|_p = 1 \\ \vect{x} \perp \vect{1}  }} \sum_{(u,v) \in E} | x_u - x_v |^p - c\left| \sum_{(u,v) \in E}  (x_u - x_v) \right|^p,
\end{equation}
	for $0 \le c \le \frac{1}{|E|^{p-1}}$. The range of $c$ is again chosen in such a way that the expression in \eqref{eq:p-spec-partition} is always non-negative, which follows from H\"older's inequality. %
Note that for $p=2$, this recovers the expression \eqref{eq:symmetry-breaking-optimisation-problem}.

\end{remark}

\begin{remark}[Directed Laplacian] \label{rem:directed-Laplacian-partition}

In a similar vein, one may also introduce an edge-alignment-incentive term to the directed Laplacian \cite{chung2005laplacians}. Assume $G$ is strongly connected and let $P$ be the transition probability matrix of a random walk on $G$. Further, let $\vect{\varphi}$ be the Perron--Frobenius eigenvector of $P$ normalised such that $\sum_{v \in V} \varphi_v = 1$. The new optimisation problem reads:
\begin{equation} \label{eq:directed-laplacian-partition}
\argmin_{\substack{ \vect{x} \in \mathbb{R}^V \\ \|\vect{x}\|_2 = 1 \\ \vect{x} \perp \sqrt{\vect{\varphi}}  }} \sum_{(u,v) \in E} \varphi_u P_{u,v} \left( \frac{x_u}{\sqrt{\varphi_u}} -  \frac{x_v}{\sqrt{\varphi_v}} \right)^2 - c \left( \sum_{(u,v) \in E} \sqrt{ \varphi_u P_{u,v} } \left( \frac{x_u}{\sqrt{\varphi_u}} -  \frac{x_v}{\sqrt{\varphi_v}} \right) \right)^2,
\end{equation}
for $0 \le c \le \frac{1}{|E|}$, where non-negativity follows again from Cauchy--Schwarz. We further point out that a variant of the correction term, where the weighting $\sqrt{\varphi_u P_{u,v}}$ of an edge is replaced by $\varphi_uP_{u,v}$, is already incorporated in the directed Laplacian itself. This is because the correction term in that case is always zero. Indeed, we have
\begin{equation}
\sum_{(u,v) \in E} \varphi_u P_{u,v} \frac{x_u}{\sqrt{\varphi_u}} = \sum_{u \in V}  x_u \sqrt{\varphi_u} \sum_{\substack{ v \in V \\ (u,v) \in E }} P_{u,v} = \sum_{u \in V} x_u \sqrt{\varphi_u} = 0
\end{equation}
and
\begin{equation}
\sum_{(u,v) \in E} \varphi_u P_{u,v} \frac{x_v}{\sqrt{\varphi_v}} = \sum_{v \in V} \frac{x_v}{\sqrt{\varphi_v}} \sum_{\substack{u \in V \\ (u,v) \in E}} \varphi_u P_{u,v} = \sum_{v \in V}  x_v \sqrt{\varphi_v} = 0.
\end{equation}
\end{remark}

\subsubsection{Main algorithm} \label{sec:bi-part-main-alg}

In order to get from the solution of the optimisation problem \eqref{eq:symmetry-breaking-optimisation-problem} to a bi-partition\footnote{Throughout the paper, we use the notation $S \sqcup T$ to denote a disjoint union.} $V = S \sqcup T$, one just takes the pointwise sign of the optimal vector. Details can be found in Algorithm \ref{alg:bi-partition}.  

\begin{algorithm}[!h]
	\DontPrintSemicolon
	\SetNlSty{textsc}{}{}
	\SetAlgoNlRelativeSize{-1}
	\caption{Direction-incentivised spectral bi-partition.\label{alg:bi-partition}}
	\KwData{A finite directed graph $G=(V,E)$.}
	\KwResult{A small cut bi-partition $S \sqcup T=V$ such that most edges between $S$ and $T$ are from $S$ to $T$.}
	\BlankLine
	$\vect{x} \leftarrow $ solution to \eqref{eq:symmetry-breaking-optimisation-problem}\;
	$S \gets \{ v \in V \mid  x_v > 0  \}$\;
	$T \gets \{ v \in V \mid  x_v \le  0  \}$\;
	\lIf{$|(S \times T) \cap E| < |(T \times S) \cap E|$}{%
		$(S, T) \leftarrow (T, S)$%
	}
	\Return $(S,T)$\;
\end{algorithm}

\subsection{Evaluation} \label{sec:eval-partition}

In this section, we shall demonstrate that Algorithm \ref{alg:bi-partition} identifies bi-partitions with more aligned cut edges than direction-oblivious algorithms. We further compare the cut quality against its classic counterpart, Equation \eqref{eq:classic-spec-partition}, and state-of-the-art (undirected) partitioners from the literature and show that our algorithm remains competitive. In other words, the direction incentive from Algorithm \ref{alg:bi-partition} only has a minor effect on the cut quality.


\subsubsection{Algorithms} \label{sec:partition-baselines}

In our evaluation, we shall be comparing our direction-incentivised spectral bi-partition, Algorithm \ref{alg:bi-partition} with $c = \frac{1}{2|E|}$, with the following bi-partitioners from the literature:
\begin{itemize}
	\item Fiedler eigenvector bi-partitioning \cite{fiedler1973algebraic, fiedler1989laplacian},
	\item METIS \cite{karypis1998fast}, and 
	\item KaHIP \cite{sandersschulz2013}.
\end{itemize}
For each algorithm which supports weight imbalance constraints, we set the maximum weight imbalance to $0.8$, cf.\@ Equation \eqref{eq:weight-imbalance}, in order to allow for a more fair comparison.

\subsubsection{Metrics} \label{sec:partition-metric}

Given a finite directed graph $G=(V,E)$ and bi-partition $U \sqcup W = V$ of its vertices, we consider the classic metrics: \emph{conductance}, \emph{cut edges}, and \emph{weight imbalance}. Furthermore, in order to measure the impact of direction incentive of our method, we measure the \emph{misaligned cut edges}.

\paragraph{Conductance} The conductance of the bi-partition is defined as

\begin{equation}
{\rm CON} = \frac{ |E_{U\times W}| + |E_{W \times U} | }{ |E_{U\times W}| + |E_{W \times U} | + \min\{ |E_{U \times U}|, |E_{W \times W} |  \} },
\end{equation}
where $E_{A\times B} = E \cap (A \times B)$ are the edges going from $A \subseteq V$ to $B \subseteq V$ in $G$.

\paragraph{Cut edges}
We compute the proportion of cut edges relative to the number of edges in order to compare across graphs:
\begin{equation}
{\rm RCE} = \frac{|E_{U\times W}| + |E_{W \times U} |}{|E|}.
\end{equation}

\paragraph{Weight imbalance}
We calculate the weight imbalance as
\begin{equation} \label{eq:weight-imbalance}
{\rm WI = \frac{||U|-|W||}{|V|}} = \frac{\max\{|U|, |W|\}}{|V|/2} - 1.
\end{equation}

\paragraph{Misaligned cut edges} We compute the proportion of misaligned cut edges relative to the number of cut edges:

\begin{equation}
{\rm RMCE} = \frac{\min\{ |E_{U\times W}|, |E_{W \times U} | \}}{|E_{U\times W}| + |E_{W \times U} |}.
\end{equation}

\subsubsection{Data sets}
\label{sec:partition-data-set}

We consider two sets of directed graphs, synthetic ones and graphs that come from a variety of real-world applications.

The synthetic ones are generated according to one of three undirected graph distributions: Erd\H{o}s--R\'enyi \cite{erdos59}, Watts--Strogatz \cite{watts98}, and stochastic block model \cite{holland83}. On each of these synthetic graphs, we artificially insert a nearly acyclic perfectly balanced bi-partition $\{A, B\}$. We do this by choosing the direction of edges as follows. For each edge which is totally within $A$, respectively $B$, we chose a direction independently and uniformly at random. For every edge between $A$ and $B$, chose the direction $B$ to $A$ with probability $\alpha$ and the direction $A$ to $B$ with probability $1-\alpha$ independently. Here, $\alpha \in [0.01, 0.1]$ is a parameter which we call the \emph{alignment probability}.

\paragraph{Erd\H{o}s--R\'enyi (ER)}
A graph with set of vertices $A \sqcup B$, where $|A|=|B|=500$, and edges inserted between any pair of vertices with independent and uniform probability $p=0.2$.

\paragraph{Watts--Strogatz (WS)} A graph with $n=1000$ vertices arranged as a circle. The sets $A$ and $B$ are chosen each as the vertices belonging to a half of the circle. Each vertex is connected to its nearest $k=50$ neighbours via an edge. For each vertex, each edge to one of its clockwise $k/2$ nearest neighbours is rewired to a uniformly random vertex independently with uniform probability $p=0.3$.

\paragraph{Stochastic Block Model (SBM)}
A graph with set of vertices $A \sqcup B$, where $|A|=|B|=500$, and edges inserted between any pair of vertices in $A$, respectively $B$, with independent and uniform probability $p_{\rm int}=0.25$. Edges for a pair of vertices $(u,v) \in A \times B$ are inserted independently with uniform probability $p_{\rm ext}=0.2$.

\paragraph{Real-world directed (RW)} We subsequently consider a set of 36 real-world directed graphs from the university of Florida sparse matrix collection~\cite{Davis11} and 4 fine-grained scheduling graphs generated by the HyperDAG\_DB~\cite{HyperDAG-DB, papp24a}. Together, these graphs emerge from a diverse set of applications, ranging from chemical process simulations to electromagnetics and fluid dynamics, and have number of vertices in the range $ [10^2, 1.6 \times 10^4]$ and number of edges in the range $ [10^3, 10^6]$. A complete list of the matrices with their associated statistics is offered in Table \ref{table:all-graphs}.

\subsubsection{Results}
\label{sec:partition-results}

\begin{figure}[!htpb]
    \centering
    \begin{minipage}{\textwidth}
        \centering
        \includegraphics[width=0.9\textwidth]{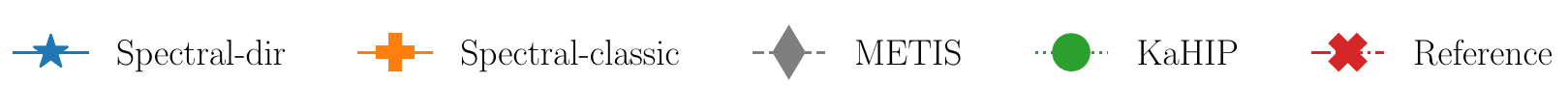}
    \end{minipage}

    \subcaptionbox{\label{fig:ER_a} Erd\H{o}s--R\'enyi (ER)}%
        {\includegraphics[width=0.225\textwidth]{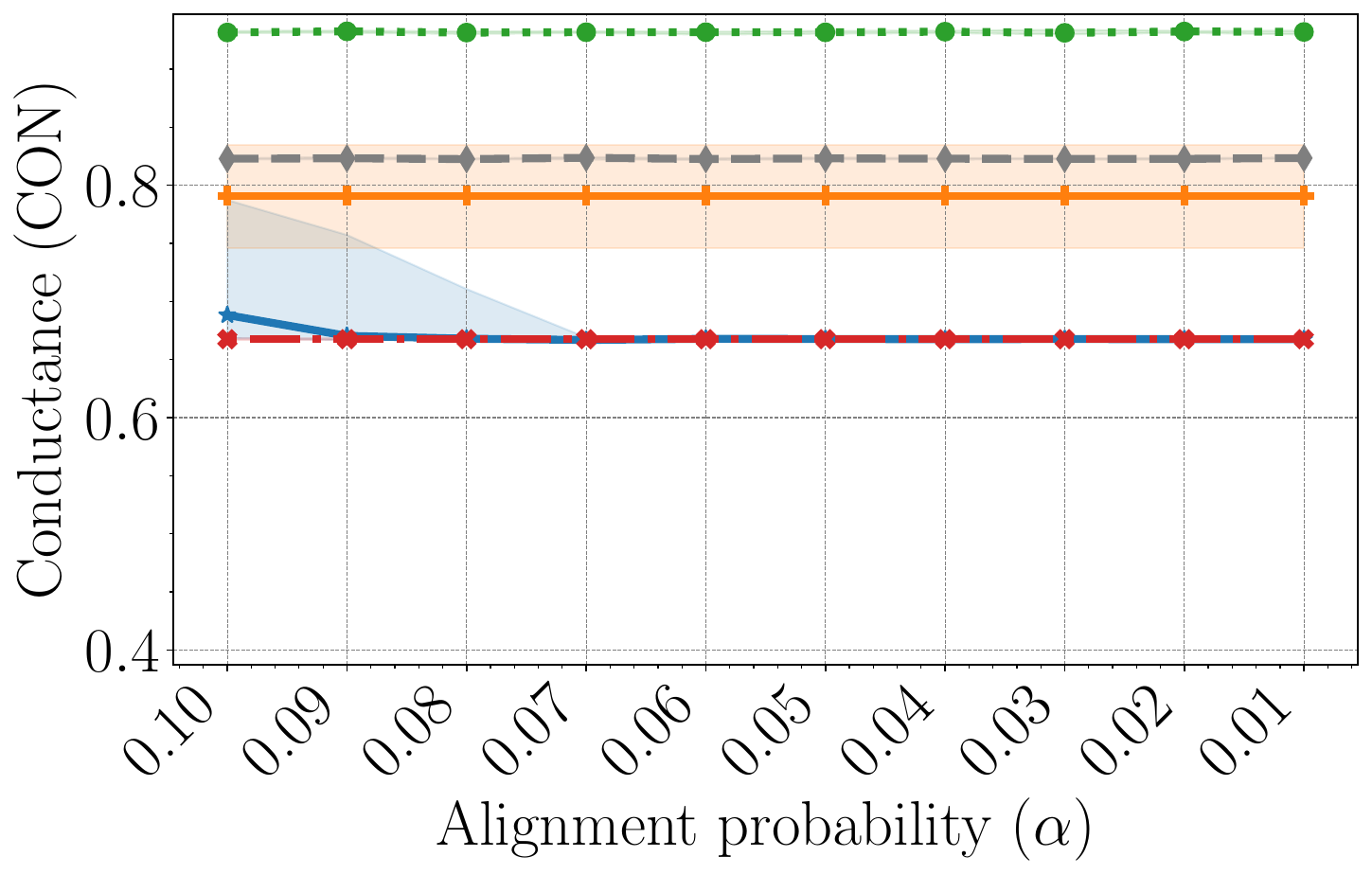}
        \includegraphics[width=0.225\textwidth]{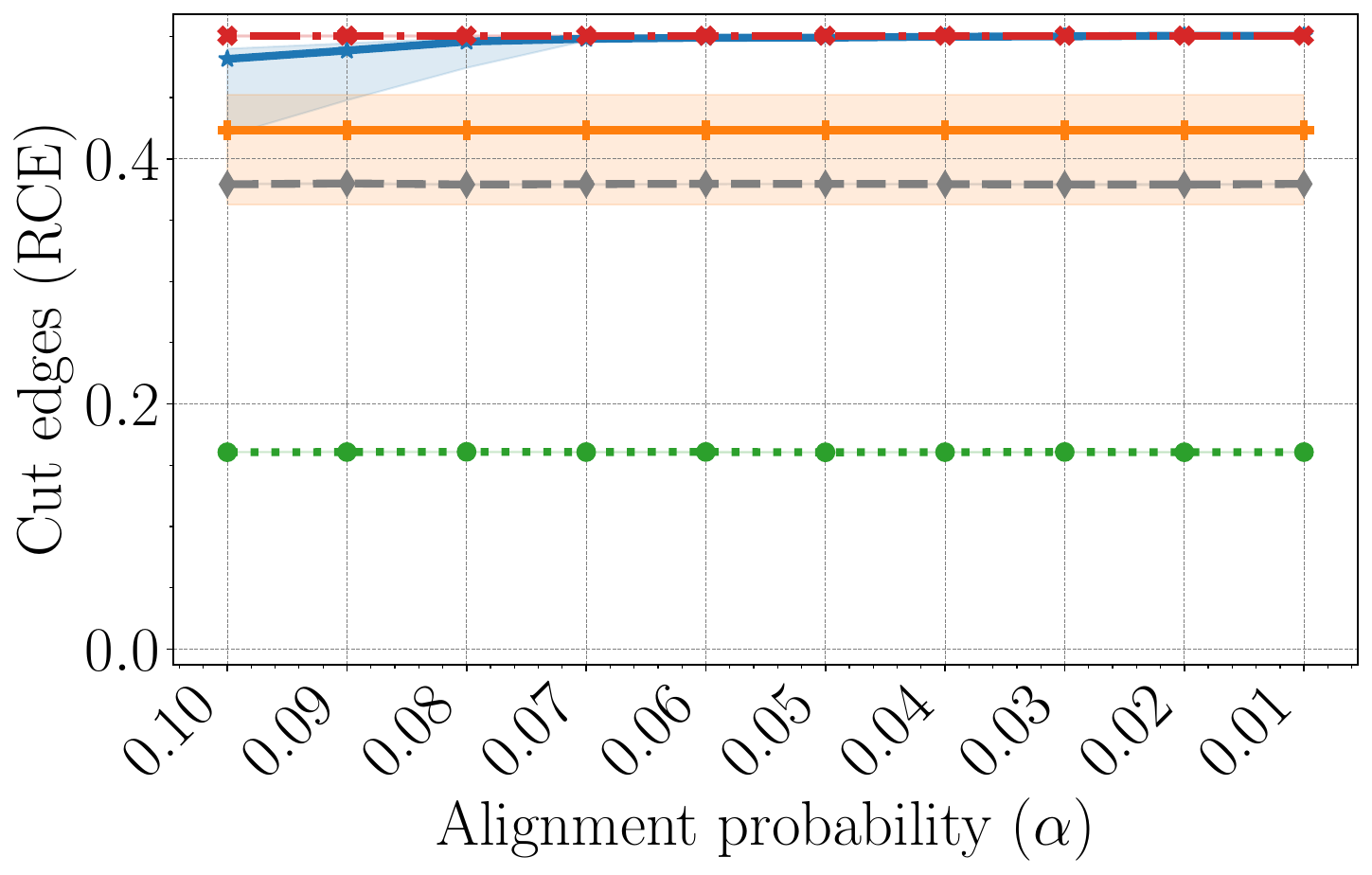}
        \includegraphics[width=0.225\textwidth]{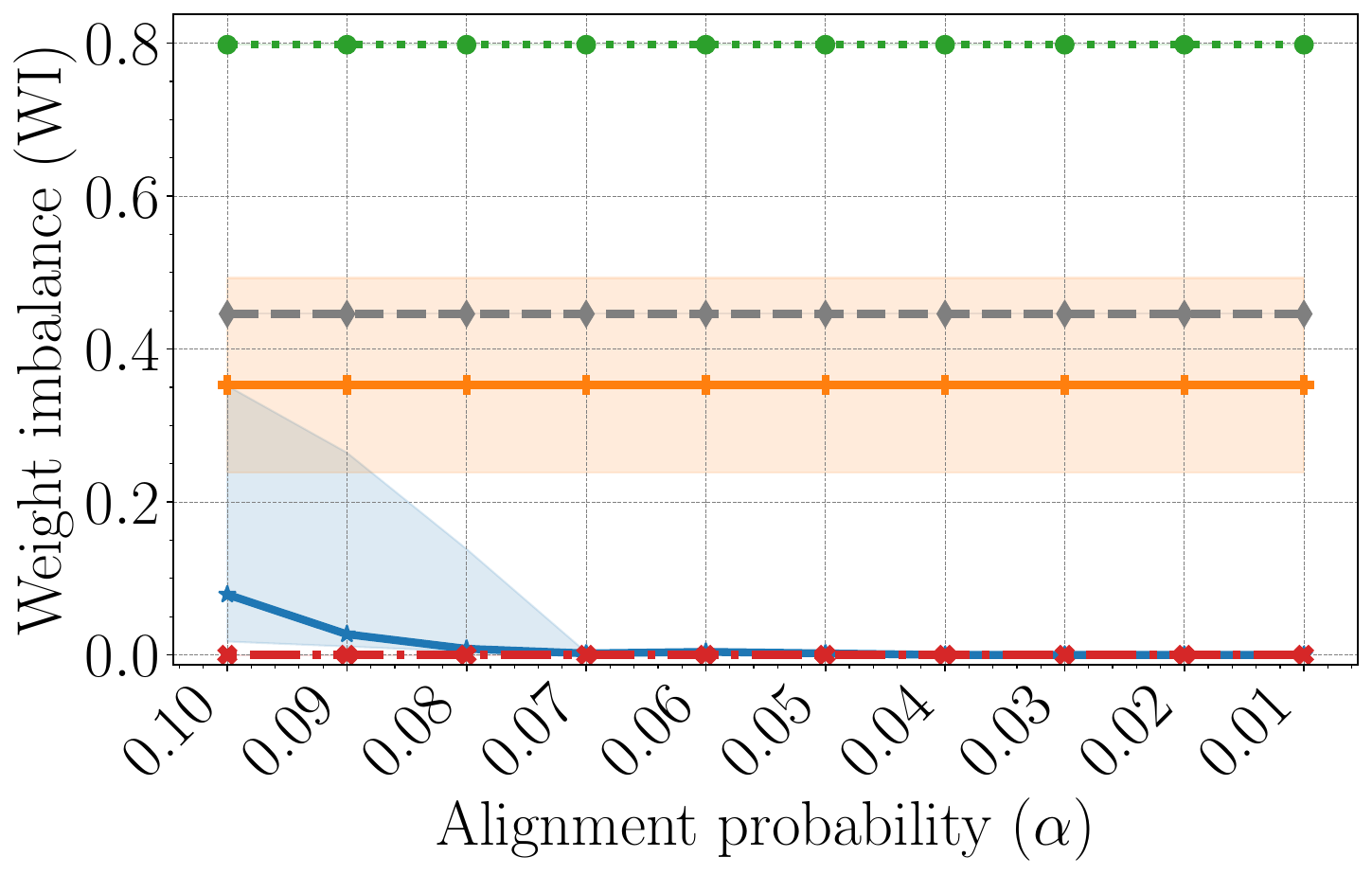}
        \includegraphics[width=0.225\textwidth]{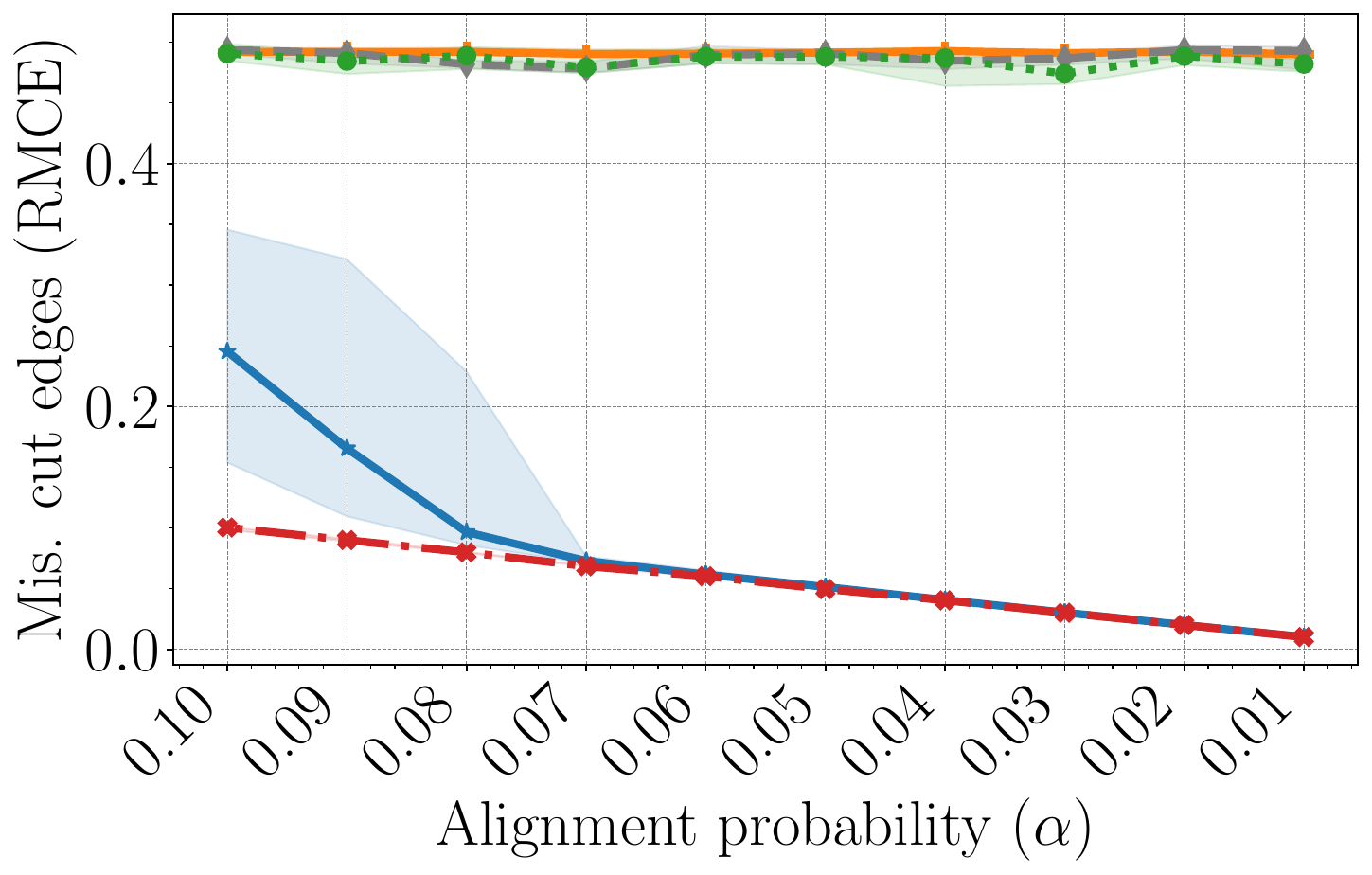}}
        \\[1em]

    \subcaptionbox{\label{fig:WS_b} Watts--Strogatz (WS)}%
        {\includegraphics[width=0.225\textwidth]{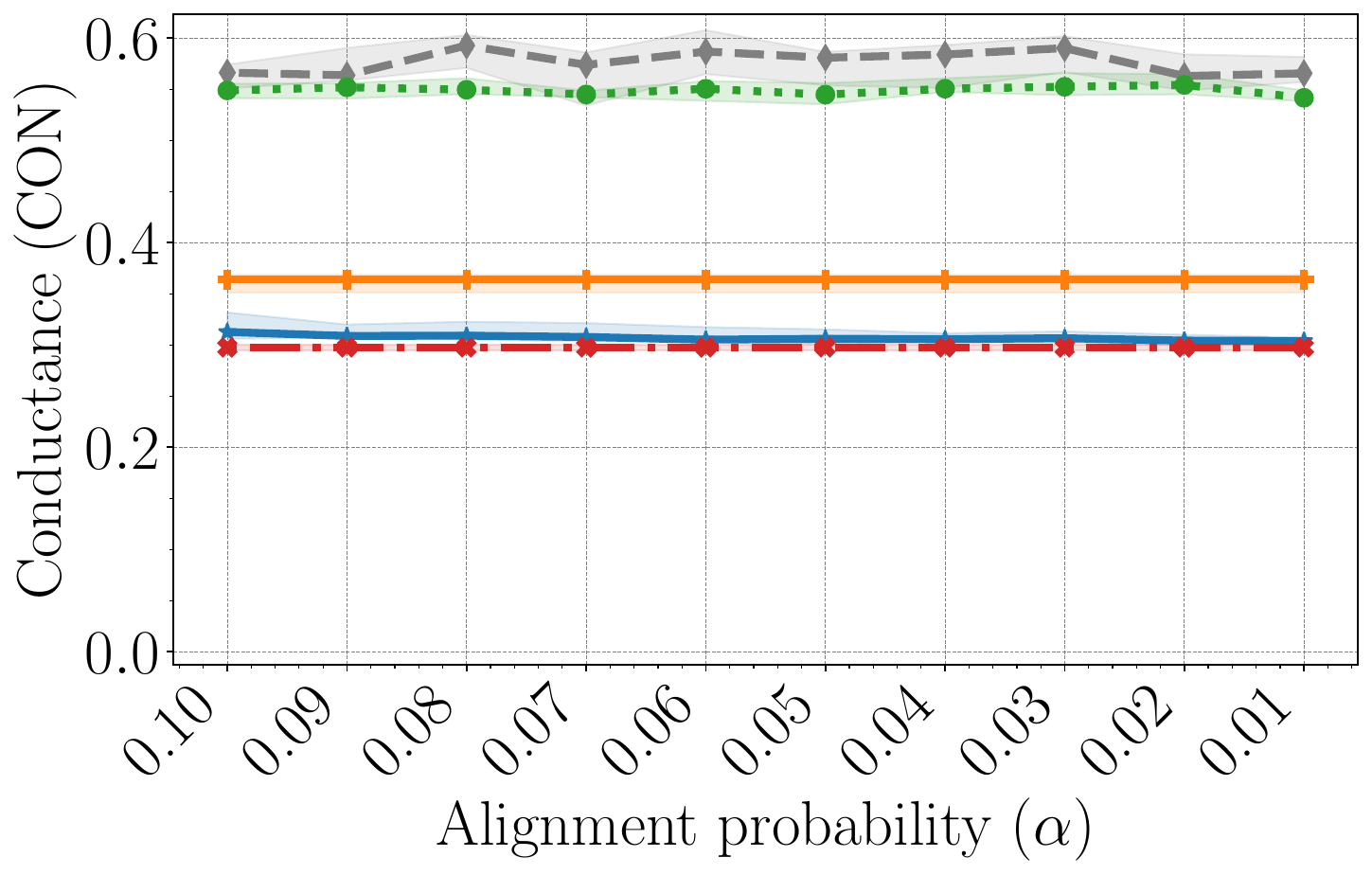}
        \includegraphics[width=0.225\textwidth]{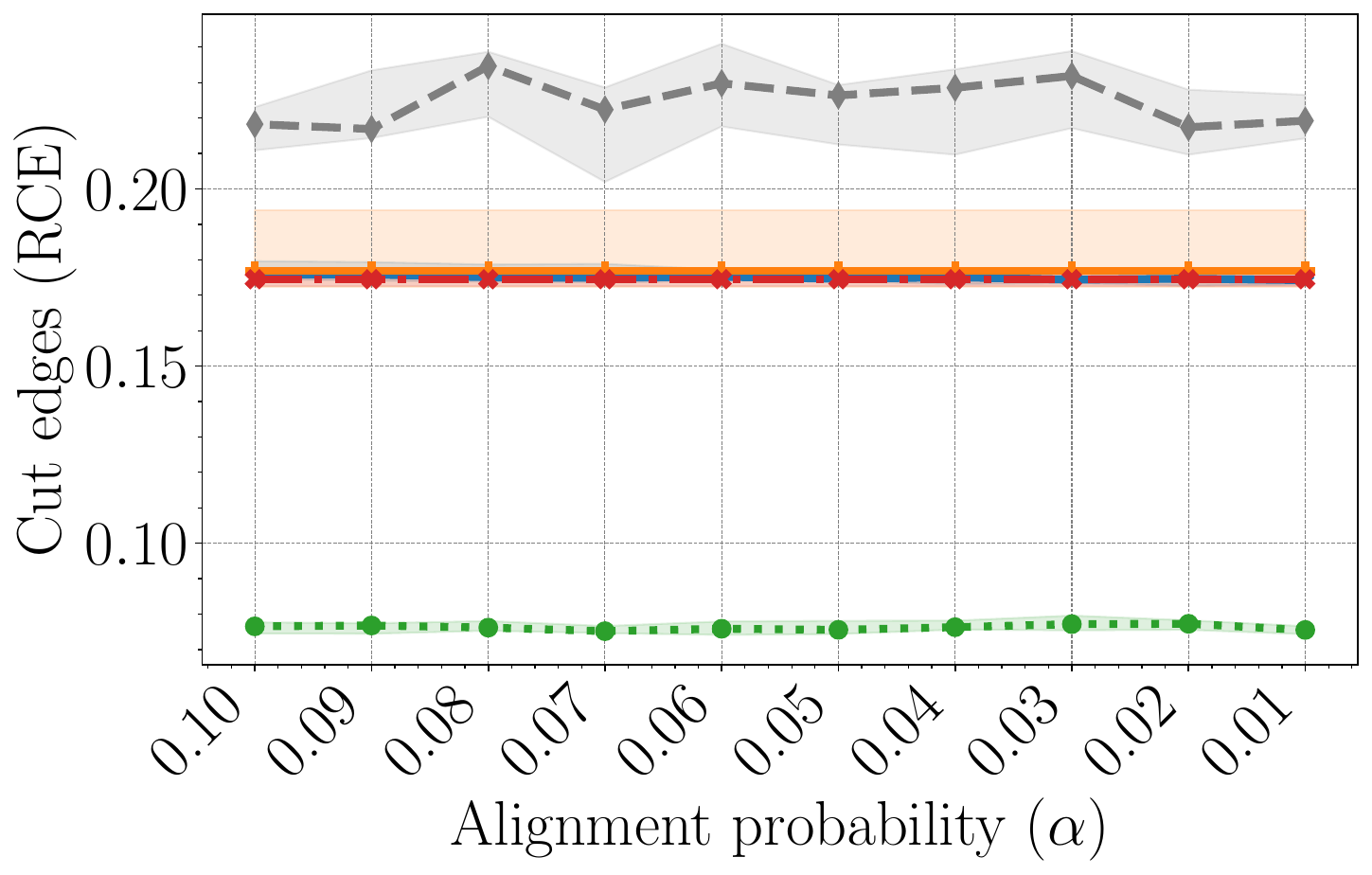}
        \includegraphics[width=0.225\textwidth]{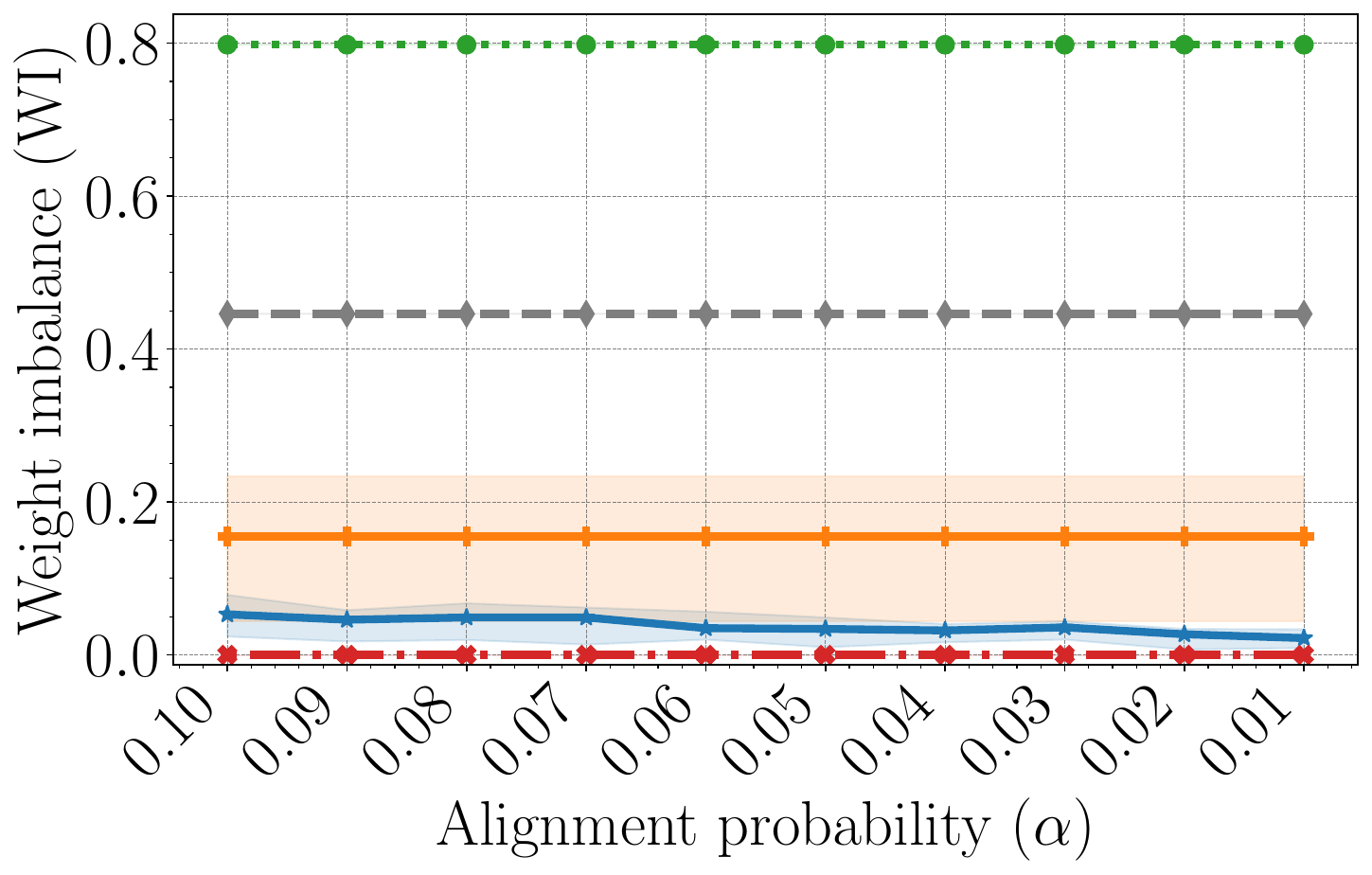}
        \includegraphics[width=0.225\textwidth]{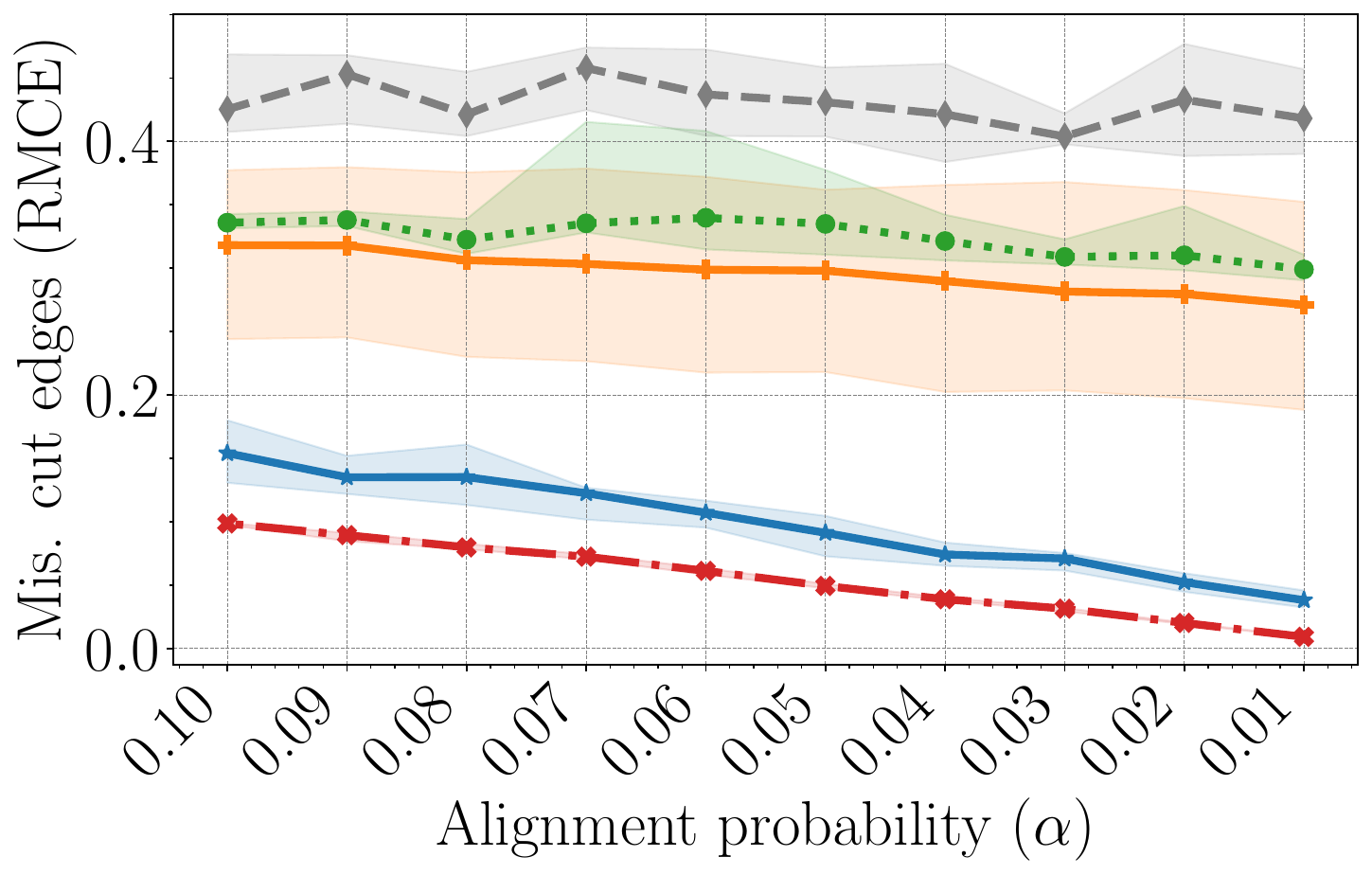}}
        \\[1em]

    \subcaptionbox{\label{fig:SBM_c} Stochastic Block Model (SBM)}%
        {\includegraphics[width=0.225\textwidth]{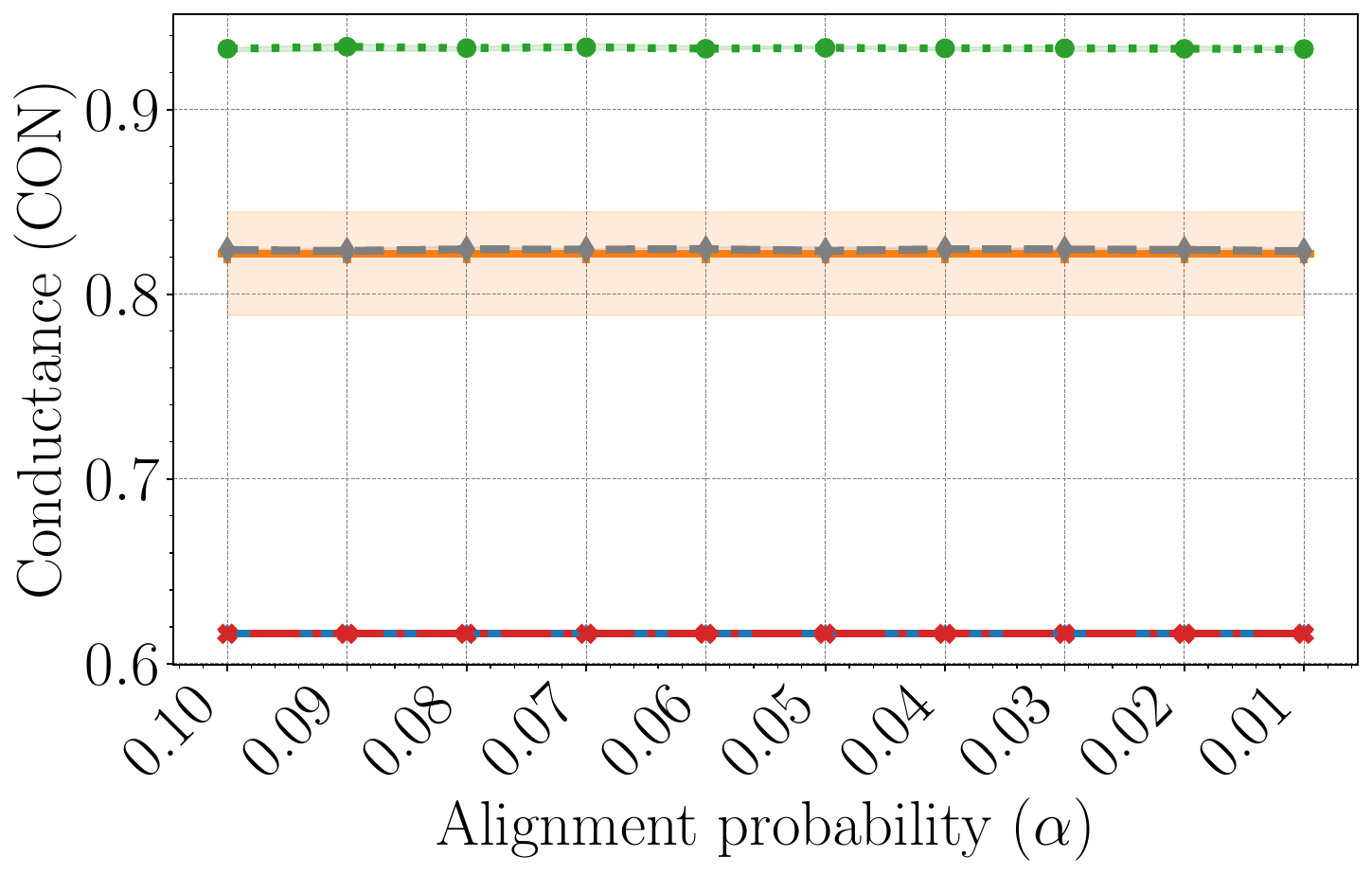}
        \includegraphics[width=0.225\textwidth]{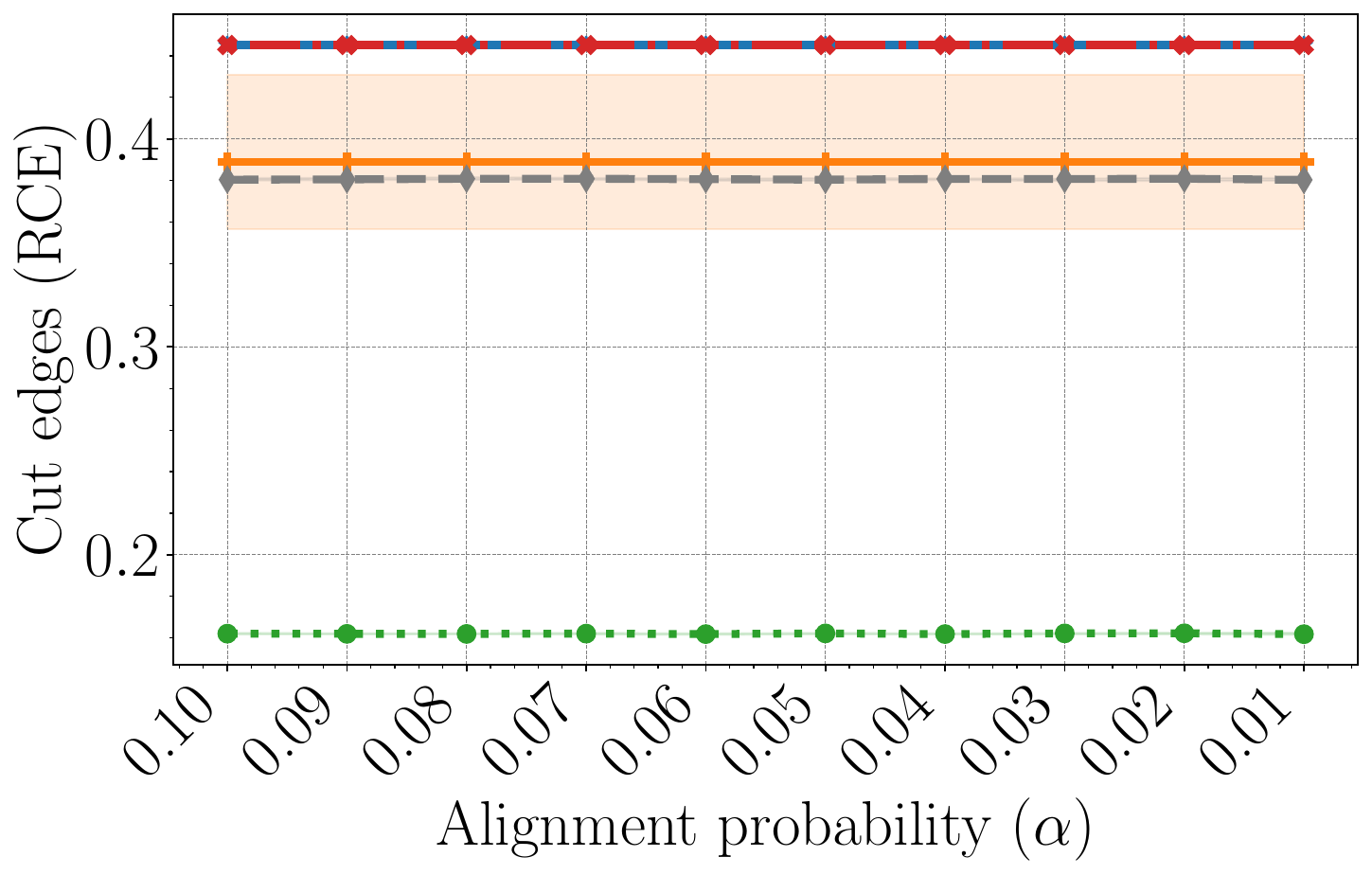}
        \includegraphics[width=0.225\textwidth]{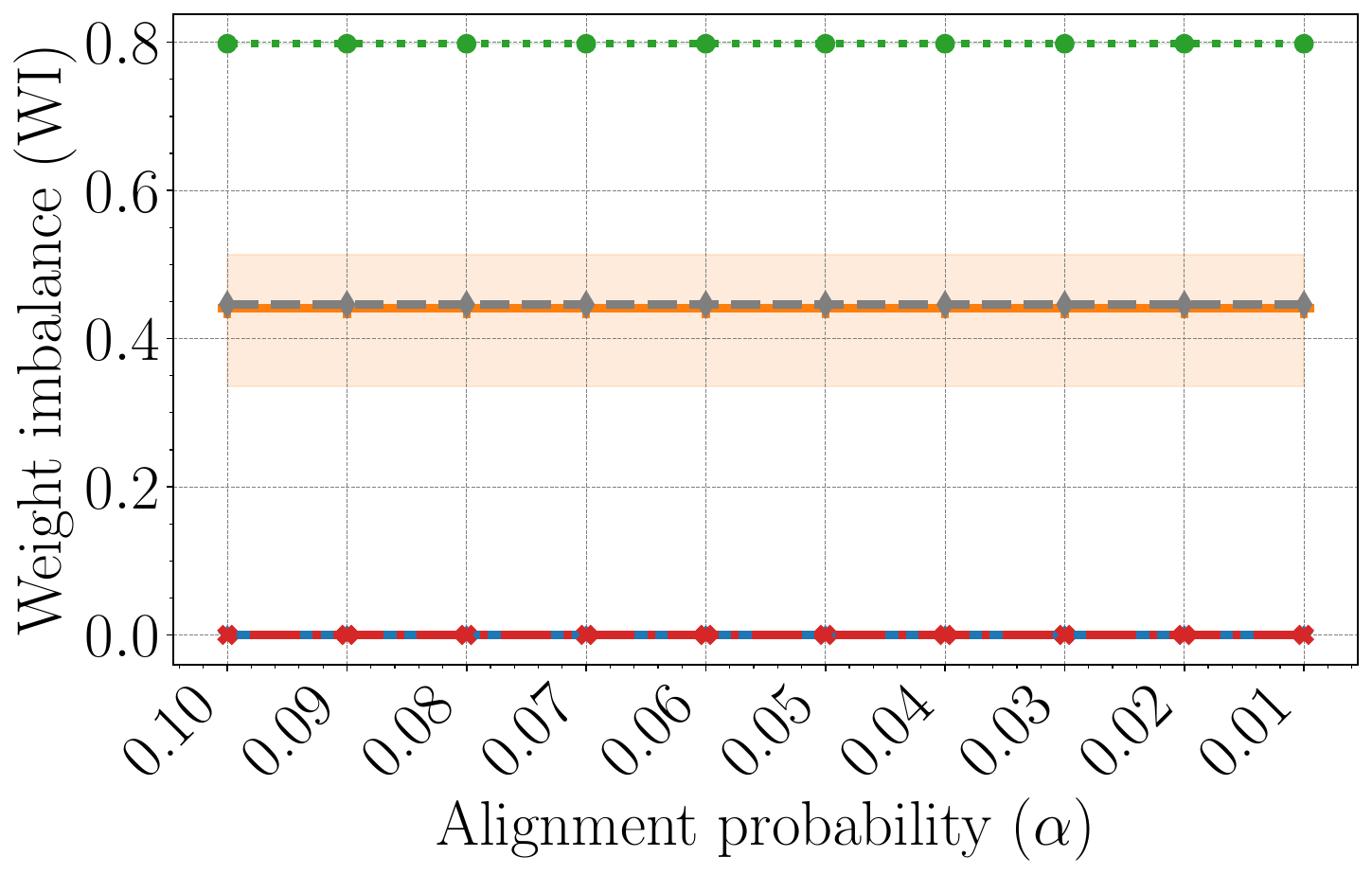}
        \includegraphics[width=0.225\textwidth]{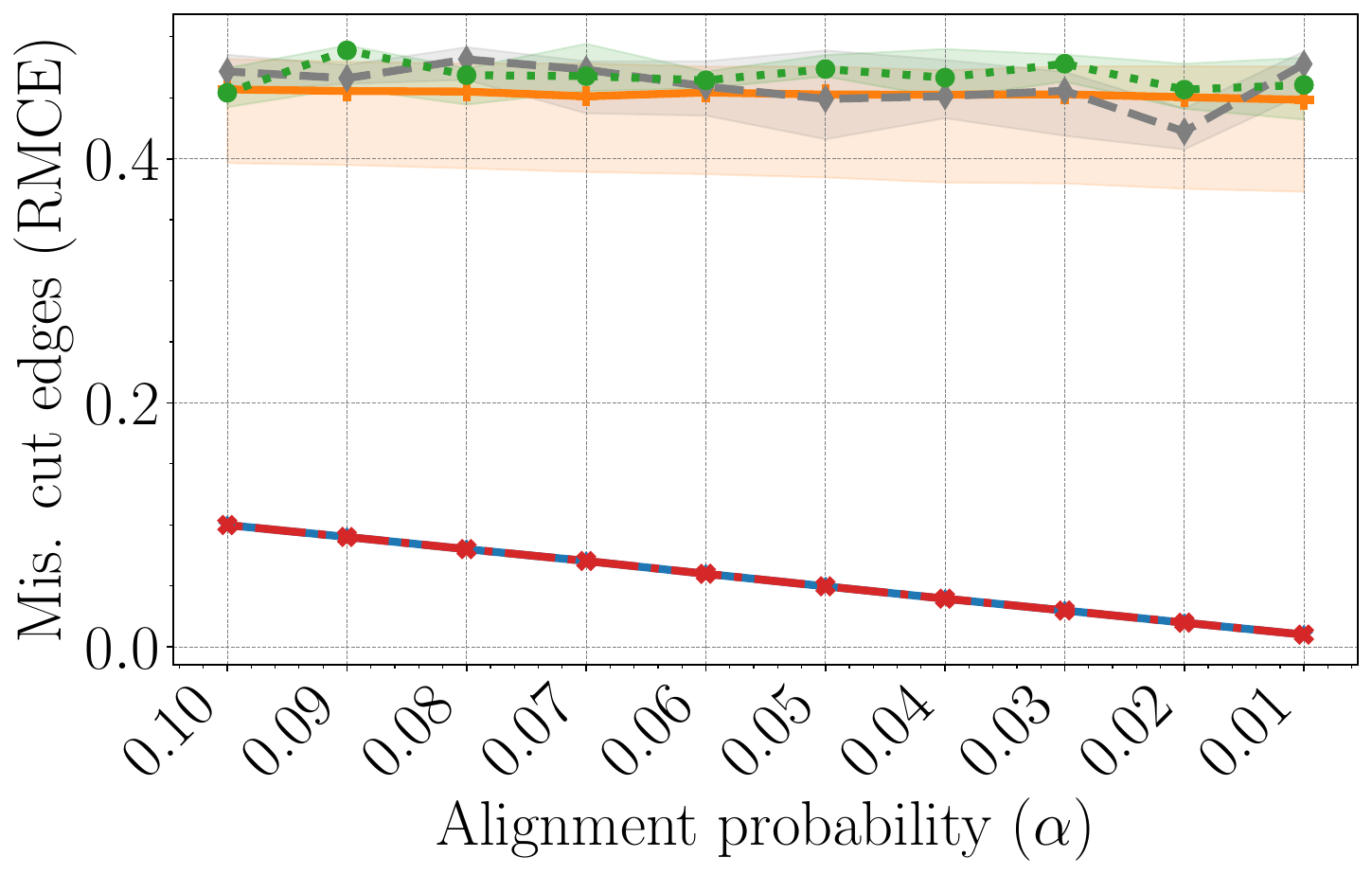}}        

    \caption{Partitioning results for conductance (CON, first column), cut edges (RCE, second column), 
    weight imbalance (WI, third column), and misaligned cut edges ratio (RMCE, right column), on the synthetic datasets of
    \S\ref{sec:partition-data-set} with a decreasing aligment
    probability $\alpha$. \ref{fig:ER_a}: Results for the Erd\H{o}s--R\'enyi (ER) graphs. \ref{fig:WS_b}: Results for the Watts--Strogatz (WS) graphs. \ref{fig:SBM_c}: Results for the Stochastic Block Model (SBM) graphs. \label{fig:Partiion_synth_res}}
\end{figure}

We present in Figure~\ref{fig:Partiion_synth_res} the bi-partitioning 
results on the synthetic ER~\ref{fig:ER_a}, WS~\ref{fig:WS_b}, and
SBM~\ref{fig:SBM_c} graphs for decreasing alignment probability 
$\alpha$. For each graph instance, we report the median and interquartile
range (IQR), i.e., the range between the $25^{\rm th}$ and $75^{\rm th}$ percentiles, for
each metric.
The performance of the four considered algorithms---our proposed 
Algorithm~\ref{alg:bi-partition} (in blue, Spectral-dir), classic spectral
bisection~\cite{fiedler1973algebraic, fiedler1989laplacian} (in orange, Spectral-classic), Metis~\cite{karypis1998fast} (in
gray), and KaHIP~\cite{sandersschulz2013} (in green)---is compared against
the ground-truth `Reference' (in red), which corresponds to the
bi-partition induced during the synthetic data generation. Perfect
alignment with the reference indicates that the algorithm accurately
recovered the induced partition.

Our method achieves conductance (CON, first column) and cut edge values (RCE, second column) closest to the reference for all three considered graph distributions.
The benefits of the direction-incentivised spectral bi-partitioning are
most apparent in the weight imbalance (WI, third column) and misaligned cut
edges (RMCE, fourth column) metrics. Spectral-dir is the only method that 
consistently identifies the induced partitions, as indicated by the proximity of WI and
RMCE values to those of the reference. Both metrics improve
(i.e., move closer to the reference) as $\alpha$ decreases, meaning fewer
cut edges are oriented against the dominant direction. For the SBM graphs,
Spectral-dir alone achieves exact correspondence with the reference across
all $\alpha$, demonstrating its ability to recover the ground-truth
partitions regardless of the cut-edge orientation.

\begin{figure}[!htb]
    \centering
    \begin{minipage}{\textwidth}
        \centering
        \includegraphics[width=0.9\textwidth]{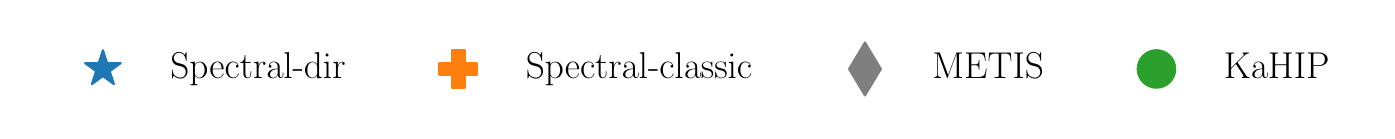}
    \end{minipage}
    \subcaptionbox{\label{fig:RW_bipart_a}}%
        {\vspace{1.6mm}\includegraphics[width=0.225\textwidth]{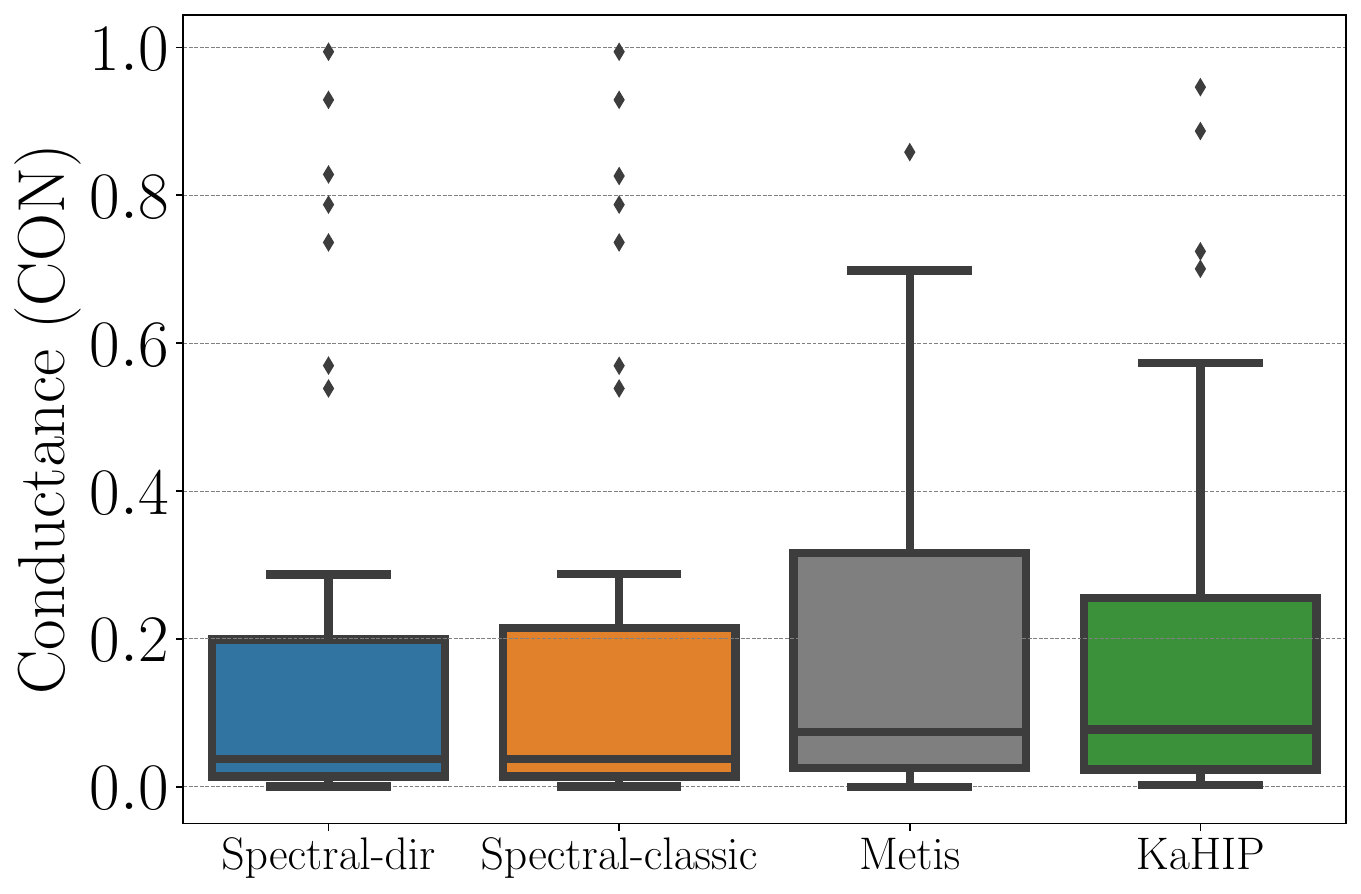}}
    \subcaptionbox{\label{fig:RW_bipart_b}}%
        {\vspace{2mm}\includegraphics[width=0.225\textwidth]{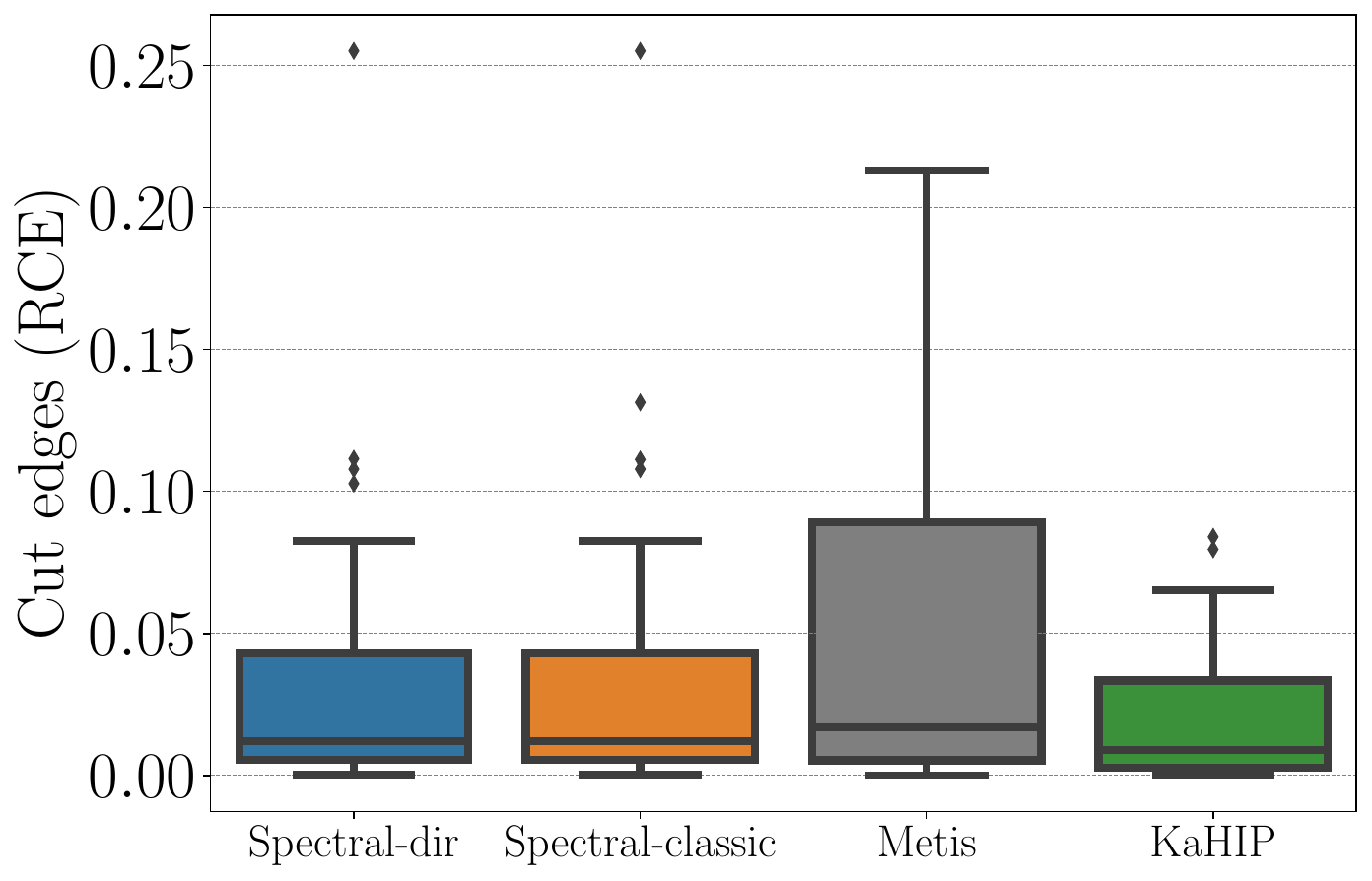}}
    \subcaptionbox{\label{fig:RW_bipart_c}}%
        {\vspace{1.6mm}\includegraphics[width=0.225\textwidth]{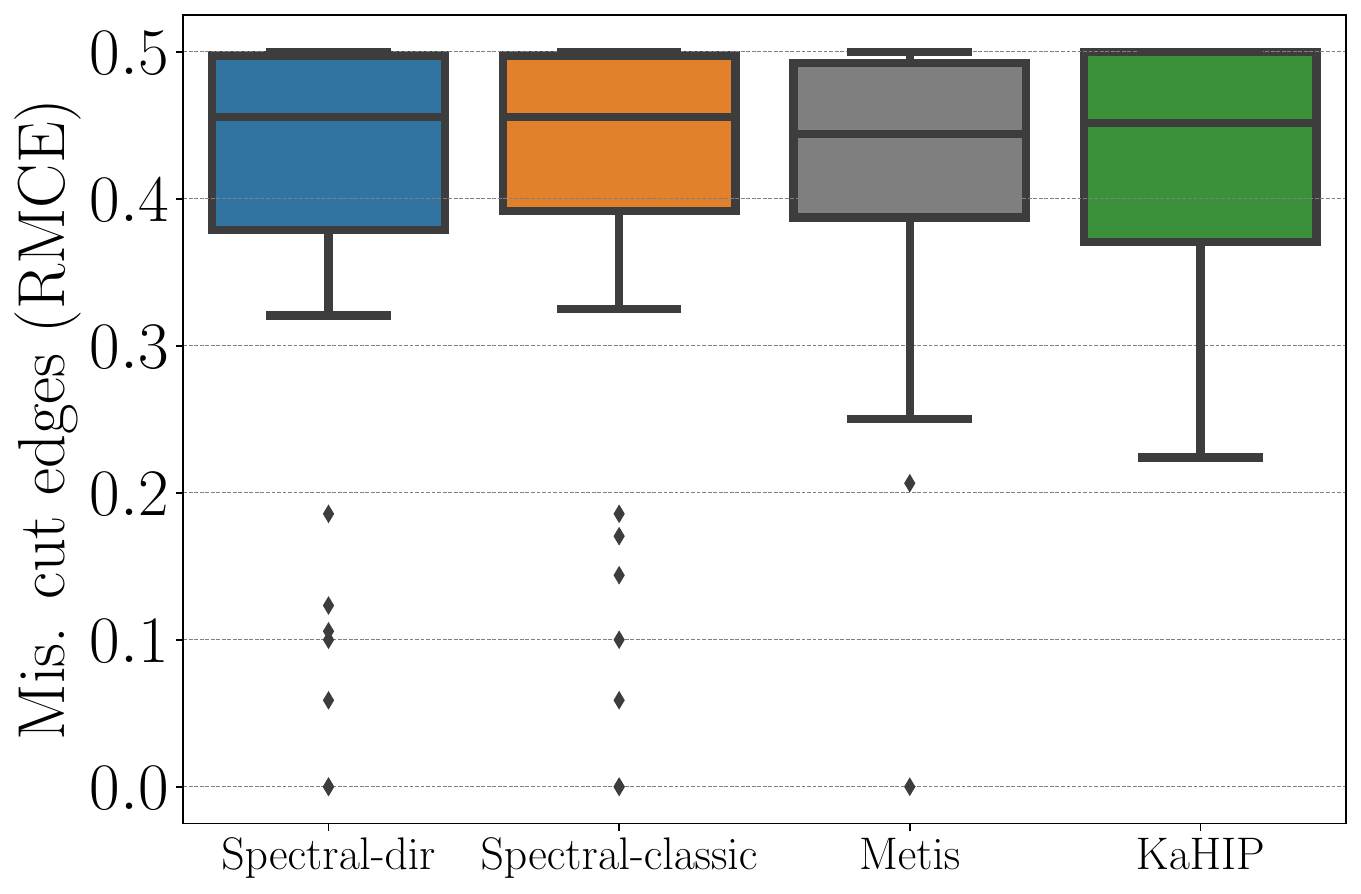}}    
    \subcaptionbox{\label{fig:RW_bipart_d}}%
        {\includegraphics[width=0.225\textwidth]{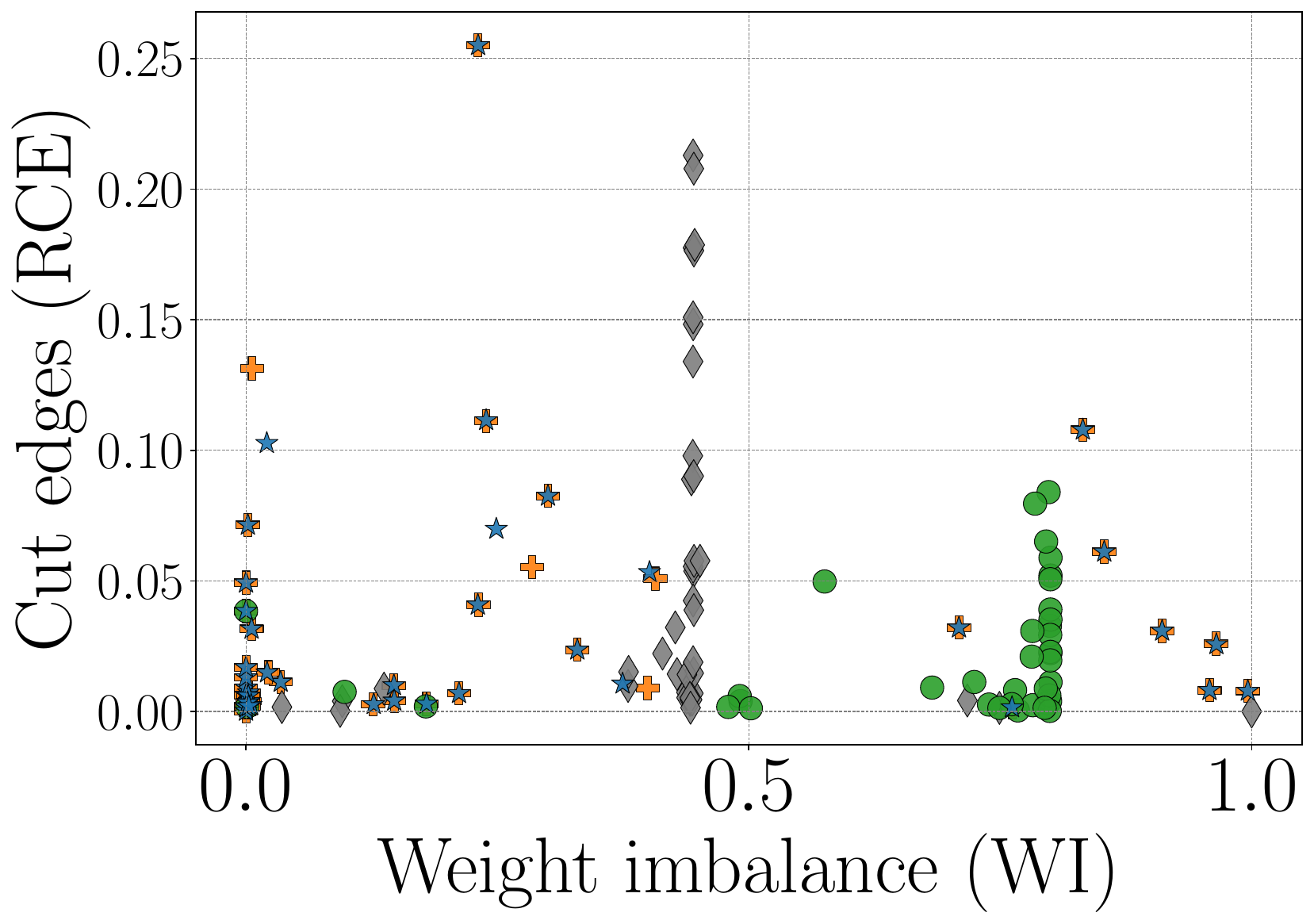}}
    \caption{\label{fig:Partiion_rw_res} Partitioning results for the real-world directed graphs of
	\S\ref{sec:partition-data-set}. \ref{fig:RW_bipart_a}: Box plot distribution of the CON results, \ref{fig:RW_bipart_b}: Box plot distribution of the RCE results, \ref{fig:RW_bipart_c}: Box plot distribution of the RMCE results, \ref{fig:RW_bipart_d}: Cut edges (RCE) over weight imbalance (WI).}
\end{figure}

For the real-world graphs, we present in Figure~\ref{fig:Partiion_rw_res}
box plots of the CON~\ref{fig:RW_bipart_a}, RCE~\ref{fig:RW_bipart_b}, and RMCE ~\ref{fig:RW_bipart_c} distribution results, and a scatter plot~\ref{fig:RW_bipart_d} to highlight the trade-off between cut edges (RCE) and weight imbalance (WI).
With respect to CON, the classic spectral method achieves the best result (strictly or tied) in 62.5\% of instances, while our proposed method does so in 55.0\%. 
Our method obtains the best (lowest) score for RMCE in 50.0\% of cases, while KaHIP reports the best results for minimizing RCE in 87.5\% of the test graphs. Our results highlight that the proposed Algorithm~\ref{alg:bi-partition} provides preferable partitioning solutions for applications where maintaining directional flow and achieving a low misalignment ratio is the primary objective.



\section{Acyclic bi-partitioning}
\label{sec:acyc-bi-partition}

In this section, we discuss a simple heuristic to turn a bi-partition into an acyclic bi-partition, whilst attempting to preserve most of the original bi-partition. 


\subsection{Algorithm}

\label{sec:acyclic-bi-partitioning-algo}

The idea we present here to rectify misaligned edges in a bi-partition of a directed acyclic graph is based on a topological order and a bisection thereof. In order to preserve as much from the original bi-partition as possible, we generate a topological order via a priority, where priority is given to vertices of the first part over those in the second part. Furthermore, as a secondary priority, vertices with a cut outgoing edge in the right direction are deprioritised whereas ones with a cut outgoing edge in the wrong direction are prioritised, and vice versa for cut incoming edges. The details can be found in Algorithm \ref{alg:acyclic-fix}, where we use the notation $\vect{1}_T$ as the indicator function of a set $T$. We point out that Herrmann \emph{et al.\@} \cite[\S4.2.2]{herrmann2019multilevel} have a similar algorithm to fix acyclicity: they move all ancestors of vertices in $S$ to $S$ or all descendants of vertices in $T$ to $T$.

\begin{algorithm}[!htpb]
	\DontPrintSemicolon
	\SetNlSty{textsc}{}{}
	\SetAlgoNlRelativeSize{-1}
	\caption{Acyclic fix.\label{alg:acyclic-fix}}
	\KwData{A finite directed acyclic graph $G=(V,E)$, a bi-partition $S \sqcup T = V$ such that $|(S\times T) \cap E| \ge |(T\times S) \cap E|$, and a balance parameter $\beta \in ]0,1[$.}
	\KwResult{A bi-partition $S' \sqcup T'= V$ such that $(T' \times S') \cap E = \emptyset$.}
	\BlankLine
	${\rm prio}[v] \leftarrow 0, \quad \forall v \in V$\;
	\For{$(u, v) \in (S \times T)\cap E$}{
		${\rm prio}[u] \leftarrow {\rm prio}[u] + 1$\;
		${\rm prio}[v] \leftarrow {\rm prio}[v] - 1$\;
	}
	\For{$(u, v) \in (T \times S)\cap E$}{
		${\rm prio}[u] \leftarrow {\rm prio}[u] - 1$\;
		${\rm prio}[v] \leftarrow {\rm prio}[v] + 1$\;
	}
	${\rm prioQ} \leftarrow \emptyset$\;
	\For{$v \in V$}{
		\lIf{\emph{number of parents of} $v \in G$ \emph{is} $0$}{%
			${\rm prioQ.insert}((\vect{1}_{T}(v), {\rm prio}[v], v))$
		}
	}
	${\rm TopOrdVec} \leftarrow \emptyset$\;
	\While{\Not ${\rm prioQ.empty}()$}{
		$(\_,\_, v) \leftarrow {\rm prioQ.popMin}()$\;
		${\rm TopOrdVec.pushback}(v)$\;
		\For{$(v, u) \in E$}{
			\lIf{\emph{all parents of} $u \in G$ \emph{are in} ${\rm TopOrdVec}$}{%
				${\rm prioQ.insert}((\vect{1}_{T}(u), {\rm prio}[u], u))$
			}
		}
	}
	$s_{\rm min} \leftarrow $ maximal $s \in \mathbb{Z}_{\ge 0}$ such that the first $s$ vertices of ${\rm TopOrdVec}$ are contained in $S$\;
	$s'_{\rm min} \gets \max\{s_{\rm min}, \min\{ |S|, \beta |V| \} \}$\;
	$t_{\rm min} \leftarrow $ maximal $t \in \mathbb{Z}_{\ge 0}$ such that the last $t$ vertices of ${\rm TopOrdVec}$ are contained in $T$\;
	$t'_{\rm min} \gets \max\{t_{\rm min}, \min\{ |T|, \beta |V| \} \}$\; 
	\Return minimal-cut bisection $(S',T')$ of ${\rm TopOrdVec}$ s.t.\@ $|S'| \ge s'_{\rm min}$ and $|T'| \ge t'_{\rm min}$\;
\end{algorithm}

\begin{remark}
	In the case where the bi-partition came from the direction-incentivised spectral partitioning, see Algorithm \ref{alg:bi-partition}, we are given more nuanced information via the solution to Equation \eqref{eq:symmetry-breaking-optimisation-problem}. This could also be used as the first priority in Algorithm \ref{alg:acyclic-fix} instead, by either using the values directly or discretising them first. Note that the solution to \eqref{eq:symmetry-breaking-optimisation-problem} may need to be multiplied with $-1$ to have the edges in the right direction.
\end{remark}

\subsection{Evaluation} \label{sec:acyc-partition-eval}

In this section, we shall demonstrate that the acyclic fix, Algorithm \ref{alg:acyclic-fix}, preserves the initial partition better when it comes from our direction-incentivised spectral partitioning method opposed to the classical spectral partitioning method.

In a second step, we show that spectral partitioning methods in conjunction with the acyclic fix, Algorithm \ref{alg:acyclic-fix}, produce acyclic bi-partitions that are on par if not better with ones produced by state-of-the-art algorithms from the literature and that this can be further improved by local-search algorithms. To this end, we compare our algorithm against ones from the literature on several graph classes using various metrics.


\subsubsection{Algorithms} \label{sec:acyc-partition-baselines}

We compare our direction-in\-cen\-tiv\-ised spectral bi-partition, Algorithm \ref{alg:bi-partition} with $c = \frac{1}{2|E|}$, together with the acyclic fix, Algorithm \ref{alg:acyclic-fix}, with the classic spectral partitioning also combined with our acyclic fix, and furthermore the following algorithms from the literature:

\begin{itemize}
	\item acyclicity-adapted Fiduccia--Mattheyses \cite{cong1994acyclic, popp2021multilevel} with initial acyclic bi-partition given by our direction-incentivised spectral bi-partition, and
	\item dagP \cite{herrmann2017acyclic, herrmann2019multilevel}.
\end{itemize}

For each algorithm which supports weight imbalance constraints, we set the maximum weight imbalance to $0.8$, cf.\@ Equation \eqref{eq:weight-imbalance}, in order to allow for a more fair comparison.

\subsubsection{Metrics} \label{sec:acyc-partition-metric}

We are given a finite directed acyclic graph $G=(V,E)$ and an acyclic bi-partition $U \sqcup W = V$ of its set of vertices. Here, acyclic means that there is no edge going from a vertex in $W$ to a vertex in $U$, that is $(W \times U)\cap E = \emptyset$.

As in \S\ref{sec:partition-metric}, we measure conductance, cut edges, and weight imbalance. In order to further evaluate the acyclicity fix, see Algorithm \ref{alg:acyclic-fix}, we measure the \emph{preserved labels} from the original bi-partition.

\paragraph{Preserved labels}

Given an original bi-partition $S \sqcup T = V$ of the set of vertices, such that $|(S \times T) \cap E | \ge | (T \times S) \cap E |$, we measure normalised number of preserved labels as:

\begin{equation}
	{\rm NPL} = \frac{|S \cap U| + |T \cap W|}{|V|}.
\end{equation}

\subsubsection{Data set} \label{sec:acyc-partition-data-set}

We consider the same set of real-world directed input graphs as in \S\ref{sec:partition-data-set}, and convert them into acyclic instances.
For an input unsymmetric adjacency matrix $A \in \mathbb{R}^{V \times V}$, we extract its strictly upper triangular part, $A_U$, and strictly lower triangular part, $A_L$. Let $G_U$ and $G_L$ be the directed graphs corresponding to $A_U$ and $A_L$, respectively.
To form the acyclic matrix $A'$, we analyse the strictly upper ($A_U$) and lower ($A_L$) triangular components of the input matrix $A$. If only one of the corresponding graphs, $G_U$ or $G_L$, is weakly connected, we select its matrix. When both are weakly connected, the selection is determined by edge density, with ties broken in favour of $A_U$. In the case where neither is weakly connected, we instead select the largest weakly connected component of whichever part, $A_U$ or $A_L$, is denser.

\subsubsection{Results} \label{sec:acyc-partition-results}

\begin{figure}[!htpb]
	\centering
	\begin{minipage}{\textwidth}
		\centering
		\includegraphics[width=0.9\textwidth]{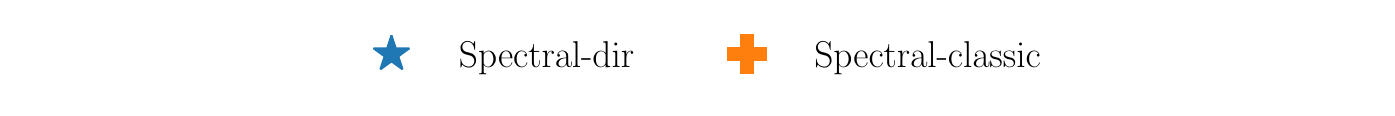}
	\end{minipage}
	\subcaptionbox{\label{fig:RW_acyc_fix_a}}%
	{\includegraphics[width=0.3\textwidth]{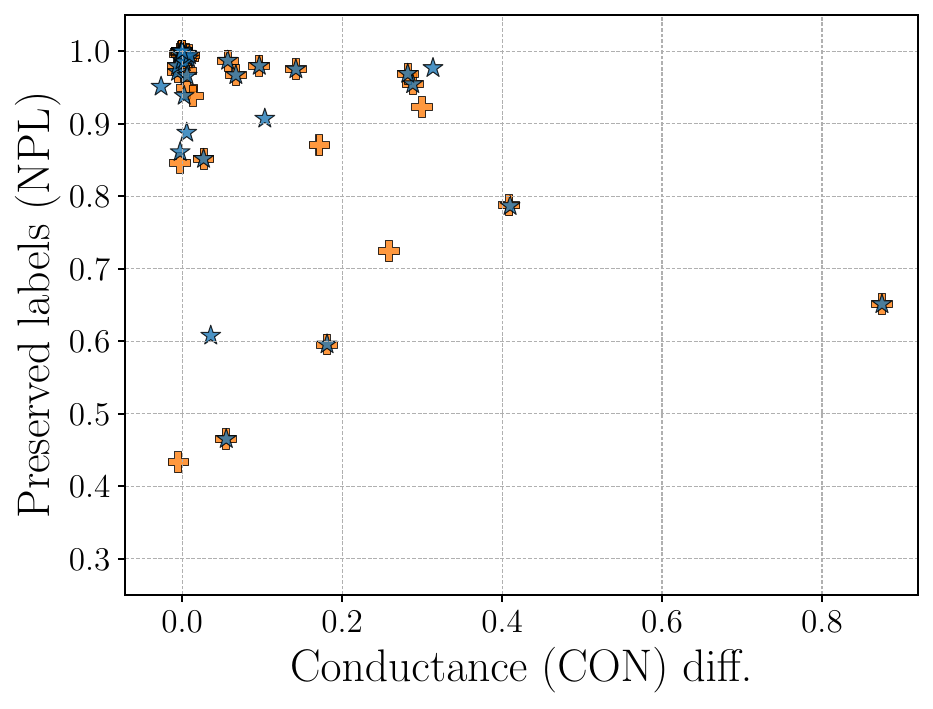}}
	\subcaptionbox{\label{fig:RW_acyc_fix_b}}%
	{\includegraphics[width=0.3\textwidth]{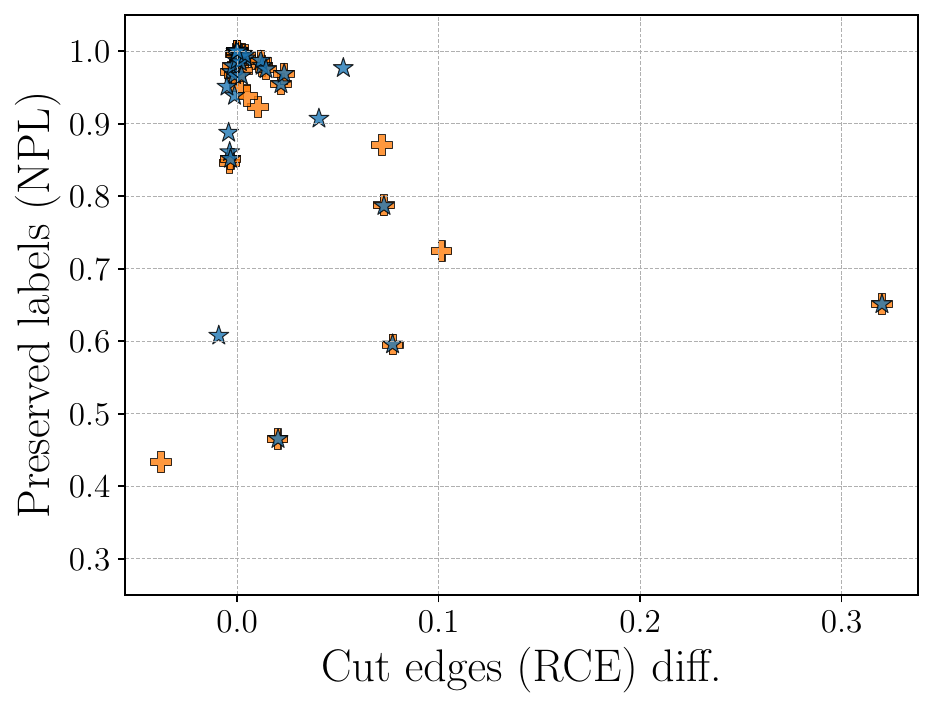}}
    \subcaptionbox{\label{fig:RW_acyc_fix_c}}%
	{\includegraphics[width=0.3\textwidth]{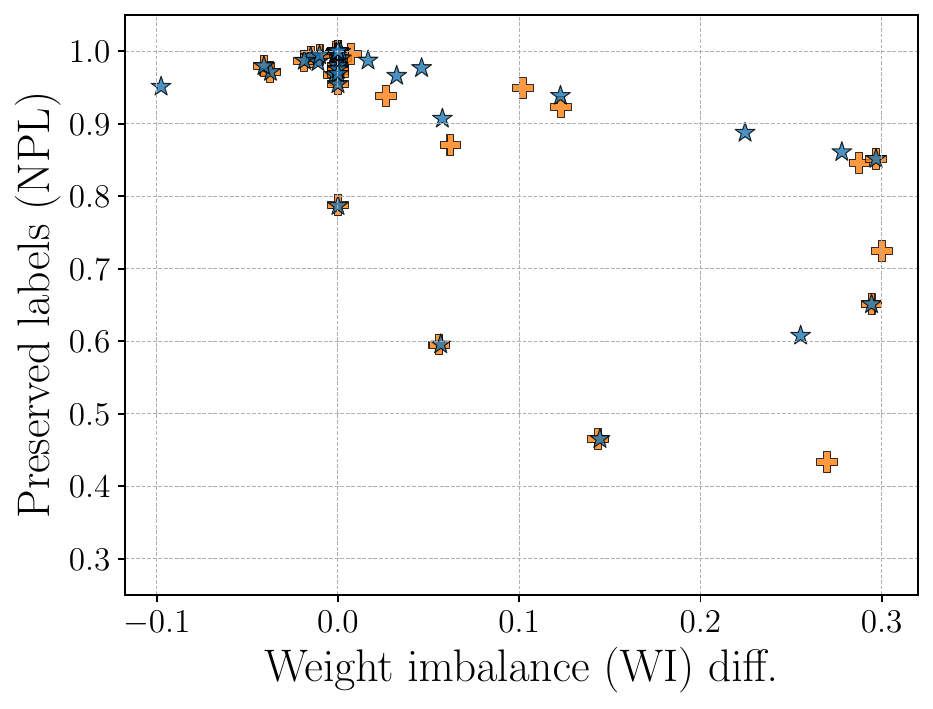}}
	\caption{\label{fig:Partiion_rw_acyc_fix}  Impact of the acyclic-fix post-processing step, introduced in Algorithm~\ref{alg:acyclic-fix}, on the spectral base bi-partitioning algorithms.
    \ref{fig:RW_acyc_fix_a}: Change in conductance (CON) over the fraction of correctly assigned vertices (NPL)
    \ref{fig:RW_acyc_fix_b}: Change in cut edges (RCE) over the fraction of correctly assigned vertices (NPL), \ref{fig:RW_acyc_fix_c}: Change in weight imbalance (WI) over the fraction of correctly assigned vertices.}
\end{figure}

In Figure~\ref{fig:Partiion_rw_acyc_fix}, we present in three scatter plots the effect
of the acyclic fix, Algorithm \ref{alg:acyclic-fix}, on the direction-incentivised spectral bi-partitioning (in blue, Spectral-dir) and on the classic variant (in orange, Spectral-classic). The classic spectral algorithm produces an inherently acyclic partition in 27.5\% of cases, compared to 20.0\% for the direction-incentivised method. However, for the partitions that do require correction, our direction-incentivised approach leverages the fix more effectively to improve the cut edges (RCE), achieving this in 43.8\% of instances. In contrast, applying the fix to the classic method is more likely to worsen the RCE (65.5\% of cases). Enforcing the acyclicity constraint results in a slight degradation in partition balance (WI) in 44.8\% of classic and 43.8\% of directed corrections, and in conductance (CON) in 72.4\% and 75.0\% of cases, respectively.
For trivial reasons, we see that a higher percentage of preserved labels (NPL) generally corresponds to smaller change in absolute difference to conductance (CON), cut edges (RCE), and weight imbalance (WI). The acyclic fix was able to preserve an average of 95.8\% of the labels over all instances when the initial bi-partition was given by Spectral-dir, compared to 93.1\% in the case of Spectral--classic.

\begin{figure}[!htpb]
	\centering
	\begin{minipage}{\textwidth}
		\centering
		\includegraphics[width=0.9\textwidth]{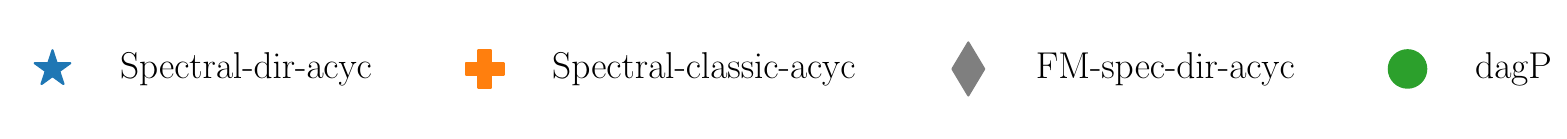}
	\end{minipage}
	\subcaptionbox{\label{fig:RW_acyc_a}}%
	{\vspace{2mm}\includegraphics[width=0.3\textwidth]{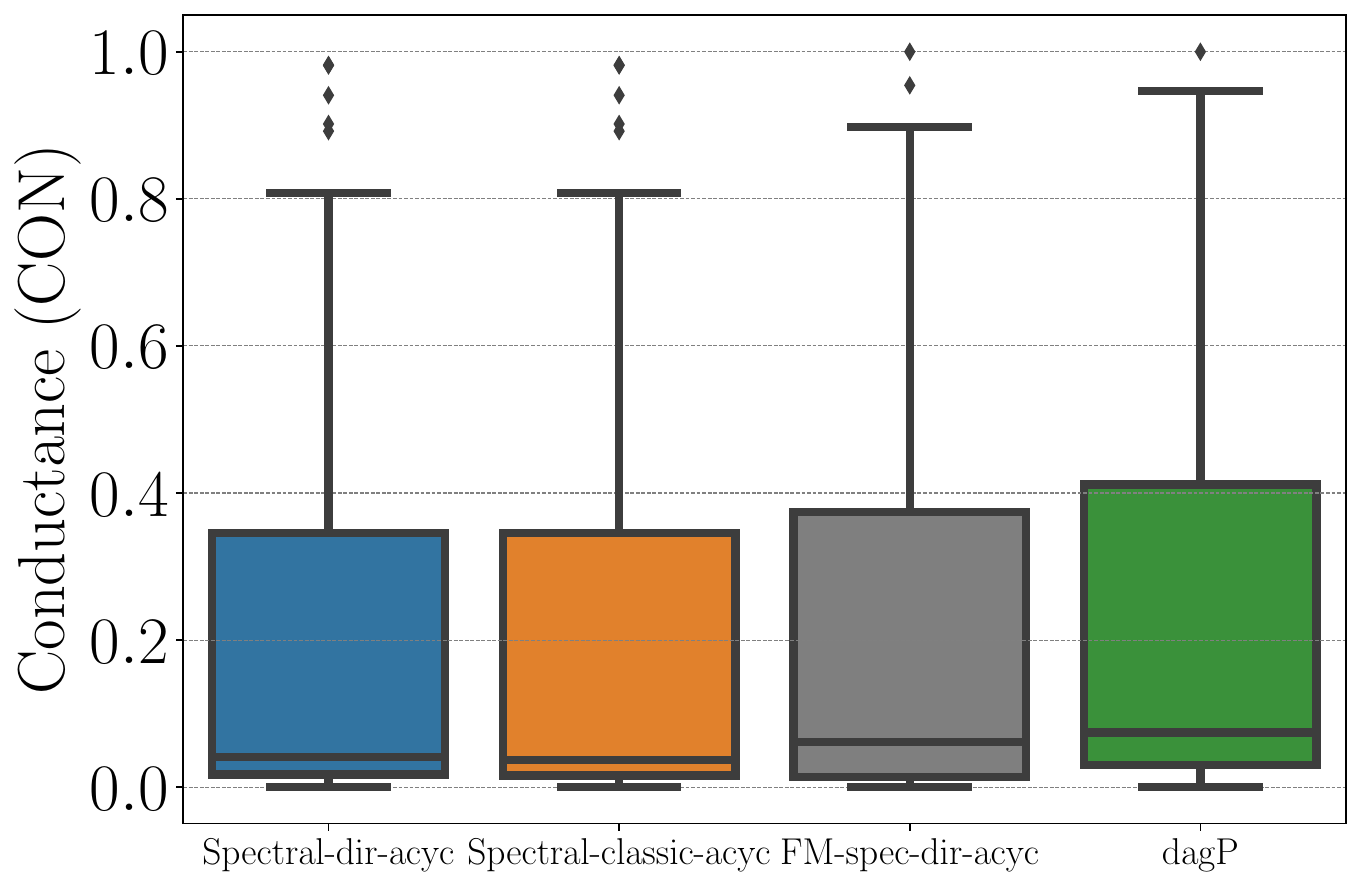}}
	\subcaptionbox{\label{fig:RW_acyc_b}}%
	{\vspace{2mm}\includegraphics[width=0.3\textwidth]{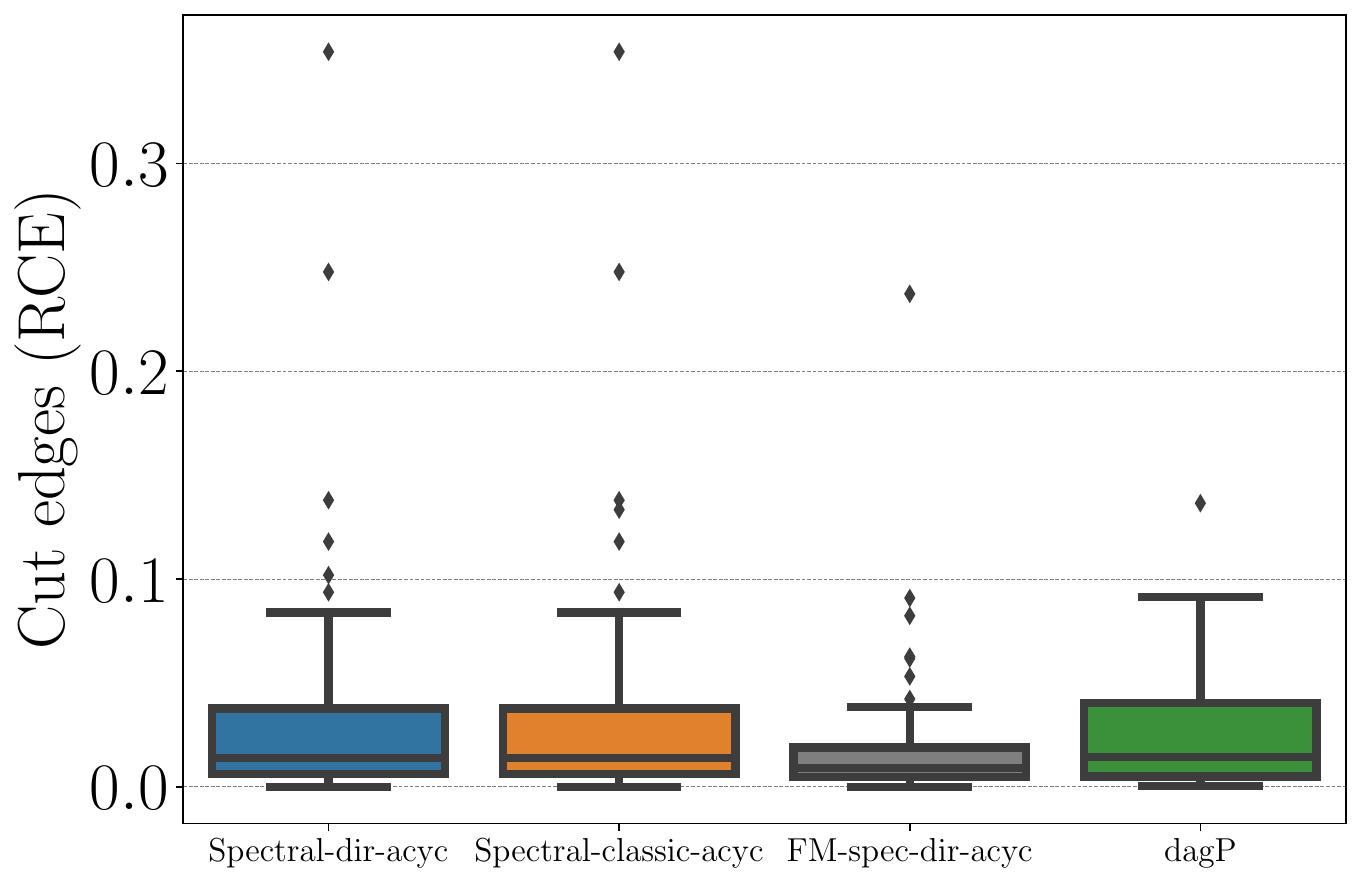}}
    \subcaptionbox{\label{fig:RW_acyc_c}}%
	{\includegraphics[width=0.3\textwidth]{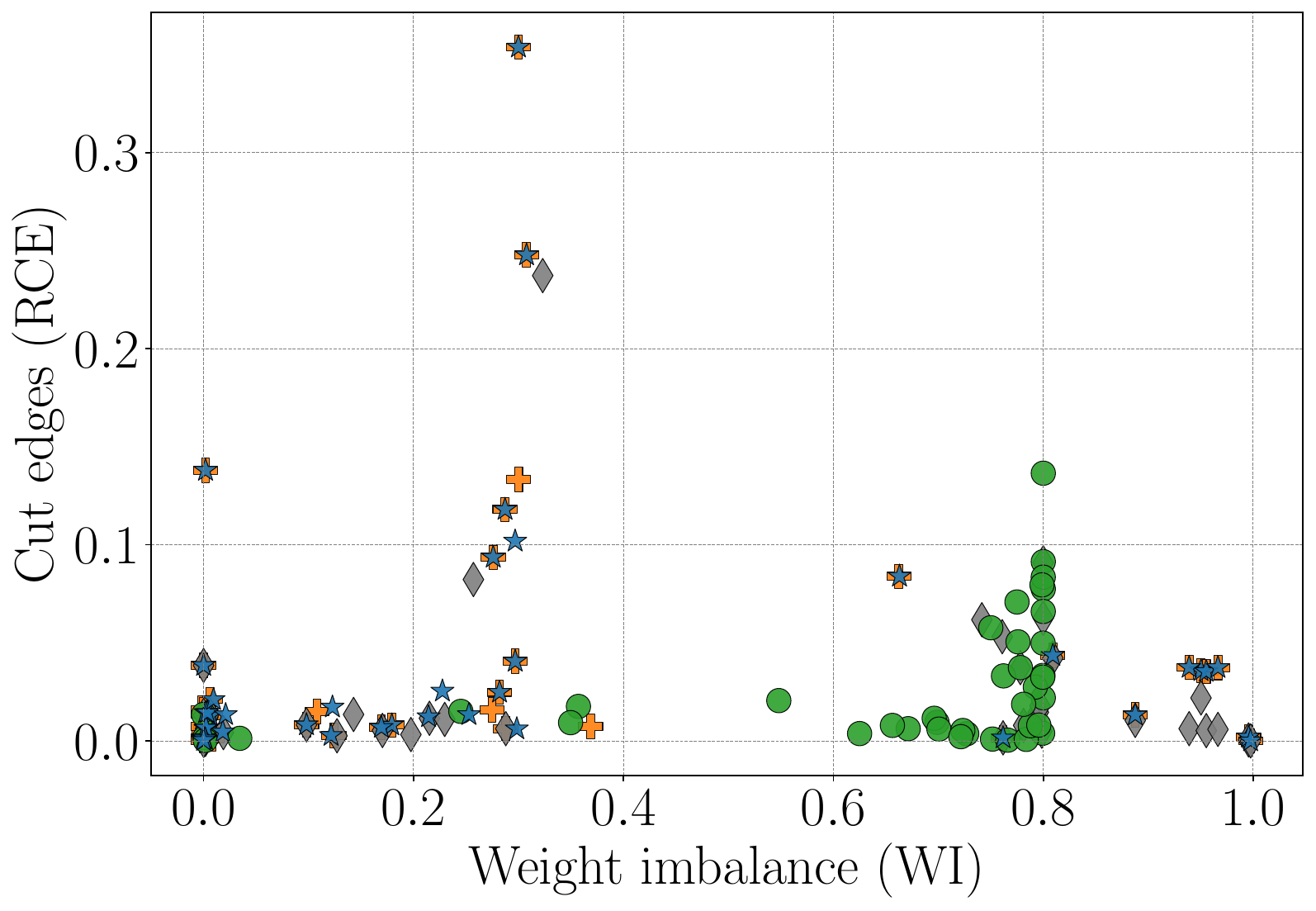}}
	\caption{\label{fig:Partiion_rw_acyc_res} Results on obtaining acyclic bi-partitions from the real-world directed acyclic graphs of \S\ref{sec:acyc-partition-data-set}.  \ref{fig:RW_acyc_a}: Box plot distribution of the CON results for each algorithm under consideration, \ref{fig:RW_acyc_b}: Box plot distribution of the RCE results for each algorithm under consideration, \ref{fig:RW_acyc_c}: Cut edges (RCE) over weight imbalance (WI).}
\end{figure}

In Figure~\ref{fig:Partiion_rw_acyc_res}, we present the bi-partitioning results on the real-world directed acyclic graphs. The performance of the four considered algorithms is shown: our proposed Algorithm~\ref{alg:bi-partition} (in blue, Spectral-dir-acyc), classic spectral bisection (in orange, Spectral-classic-acyc), the acyclicity-adapted Fiduccia-Mattheyses variant using Algorithm~\ref{alg:bi-partition} as initialiser (in gray, FM-spec-dir-acyc), and dagP (in green). The first three methods incorporate the acyclic fix introduced in Algorithm~\ref{alg:acyclic-fix}.
In Subfigures~\ref{fig:RW_acyc_a},~\ref{fig:RW_acyc_b}
we show box plots of the CON and RCE distribution results
respectively, and in~\ref{fig:RW_acyc_a} the trade off
between RCE and WI as a scatter plot.
For CON, the two spectral methods exhibit the
lowest median values, with the FM-based variant and classic spectral most frequently achieving the best score (47.5\% each). Regarding RCE, the FM-based method achieves the lowest median, with dagP obtaining the best score in the highest number of instances (60.0\%). The scatter plot in Subfigure~\ref{fig:RW_acyc_c} illustrates that the performance of dagP on RCE is associated with high weight imbalance (WI), a metric for which the classic spectral method attains the best result in 57.5\% of cases, followed by the direction-incentivized proposed variant in 50.0\%. These results highlight the effectiveness of the
FM-based refinement of spectral methods to retrieve bi-partitions with low conductance and number of cut edges.

\section{Spectral topological order} \label{sec:spec-top-order}

\subsection{Algorithm} \label{sec:alg-spec-top-order}

In this section, we describe how the cut-edge direction-incentivised spectral bi-partitioning from \S\ref{sec:bi-part-alg} can be used to generate a topological order of a directed acyclic graph.

At its core, the bi-partition given by Algorithm \ref{alg:bi-partition} is used to separate the vertices into ones that come first in a topological order and ones that come later. On each of those parts, the algorithm is applied recursively. This results in an order of the vertices which is not necessarily a topological order. In order to remedy this, we apply a direction fix after each bi-partition $S \sqcup T = V$ such that cut edges only originate from the first part $S$, that is $(T \times S) \cap E = \emptyset$. Our scheme resembles the one by Schamberger--Wierum \cite{schamberger2004locality} for undirected graphs, but with the added difficulty of generating a linear layout which is also a topological order.

In \S \ref{sec:fix-spec-top-order}, we describe the direction-fix algorithm and in \S\ref{sec:main-alg-spec-top-order}, we describe the complete topological-order algorithm.

\subsubsection{Direction fix} \label{sec:fix-spec-top-order}
In order to state the algorithm, we require some notation. Given a directed acyclic graph $G=(V,E)$ and an (acyclic) partition $\bigsqcup_{i=0}^{\ell} L_i = V$ of its set of vertices, we write
\begin{equation} \label{eq:def-vert-precedence-list}
L_1 \prec L_2 \prec \dots \prec L_{\ell}
\end{equation}
if and only if there are no edges originating from $L_j$ and ending in $L_i$ for all $j>i$. In other words, this is a topological order on the coarsened graph along the partition.

We present the direction-fix algorithm in Algorithm \ref{alg:direction-fix}, where we use the notation $\vect{1}_T$ as the indicator function of a set $T$.

\begin{algorithm}[!htpb]
	\DontPrintSemicolon
	\SetNlSty{textsc}{}{}
	\SetAlgoNlRelativeSize{-1}
	\caption{Direction fix.\label{alg:direction-fix}}
	\KwData{A finite directed acyclic graph $G=(V,E)$, a partition $K \sqcup L \sqcup M = V$ such that $K \prec L \prec M$, and a bi-partition $S \sqcup T = L$.}
	\KwResult{A bi-partition $S' \sqcup T'= L$ such that $(T' \times S') \cap E = \emptyset$.}
	\BlankLine
	${\rm prio}[v] \leftarrow 0, \quad \forall v \in L$\;
	\For{$v \in L$}{
		${\rm prio}[v] \leftarrow {\rm prio}[v] + |(\{v\}\times M)\cap E|$\;
		${\rm prio}[v] \leftarrow {\rm prio}[v] - |(K \times \{v\})\cap E|$\;
	}
	\For{$(u, v) \in (S \times T)\cap E$}{
		${\rm prio}[u] \leftarrow {\rm prio}[u] + 1$\;
		${\rm prio}[v] \leftarrow {\rm prio}[v] - 1$\;
	}
	\For{$(u, v) \in (T \times S)\cap E$}{
		${\rm prio}[u] \leftarrow {\rm prio}[u] - 1$\;
		${\rm prio}[v] \leftarrow {\rm prio}[v] + 1$\;
	}
	${\rm prioQ} \leftarrow \emptyset$\;
	\For{$v \in L$}{
		\lIf{\emph{number of parents of} $v \in G|_L$ \emph{is} $0$}{%
			${\rm prioQ.insert}((\vect{1}_{T}(v), {\rm prio}[v], v))$
		}
	}
	${\rm TopOrdVec} \leftarrow \emptyset$\;
	\While{\Not ${\rm prioQ.empty}()$}{
		$(\_,\_, v) \leftarrow {\rm prioQ.popMin}()$\;
		${\rm TopOrdVec.pushback}(v)$\;
		\For{$(v, u) \in (L \times L) \cap E$}{
			\lIf{\emph{all parents of} $u \in G|_L$ \emph{are in} ${\rm TopOrdVec}$}{%
				${\rm prioQ.insert}((\vect{1}_{T}(u), {\rm prio}[u], u))$
			}
		}
	}
	$S' \leftarrow$ first $|S|$ vertices of ${\rm TopOrdVec}$\;
	$T' \leftarrow$ last $|T|$ vertices of ${\rm TopOrdVec}$\;
	\Return $(S',T')$\;
\end{algorithm}

Algorithm~\ref{alg:direction-fix} shares a lot of similarities with Algorithm~\ref{alg:acyclic-fix}. There are, however, two important differences. First, the secondary priority also takes into account the vertices which come before and after in the vertex precedence list ${\rm VertPrecList}$. Second, the bisection of the topological order is done in a manner which preserves the cardinality of the initial bi-partition. The former change is an optimisation to keep cut edges short and the latter change is to preserve more of the spectral information.

\subsubsection{Main algorithm} \label{sec:main-alg-spec-top-order}

The main algorithm operates on a vertex precedence list ${\rm VertPrecList}$:
\begin{equation}
L_1 \prec L_2 \prec \dots \prec L_{\ell}, 
\end{equation}
where $\bigsqcup_{i=0}^{\ell} L_i = V$ is a partition of the set of vertices $V$. This list is initialised simply as $V$. In each step, a set of vertices $L_i$ with $|L_i| \ge 2$ is chosen and refined as $S \prec T$, that is, $L_i = S \sqcup T$ with $S,T \neq \emptyset$ and $E \cap (T \times S) = \emptyset$. The vertex precedence list is subsequently updated as
\begin{equation}
\cdots \prec L_{i-1} \prec S \prec T \prec L_{i+1} \prec \cdots \ .
\end{equation}
One easily verifies that this is indeed a valid vertex precedence list in the sense of Equation~\eqref{eq:def-vert-precedence-list}. The algorithm terminates as soon as all of the sets $L_i$ consist of a single vertex. The details of the algorithm may be found in Algorithm \ref{alg:spec-top-order-main}.

\begin{algorithm}[!htbp]
	\DontPrintSemicolon
	\SetNlSty{textsc}{}{}
	\SetAlgoNlRelativeSize{-1}
	\caption{Spectral topological order.\label{alg:spec-top-order-main}}
	\KwData{A finite directed acyclic graph $G=(V,E)$.}
	\KwResult{A topological order of $G$.}
	\BlankLine
	${\rm VertPrecList} \leftarrow \{V\}$\;
	\While{$\exists L \in {\rm VertPrecList}$ \emph{with} $|L| > 1$}{
		Let $L \in {\rm VertPrecList}$ with $|L|> 1$\;
		Let $K$ be the union of all sets $N \prec L$ in ${\rm VertPrecList}$ \label{alg-line:main-top-begin-acyc-split}\;
		Let $M$ be the union of all sets $N \succ L$ in ${\rm VertPrecList}$\;
		$\vect{x} \leftarrow $ solution to \eqref{eq:symmetry-breaking-optimisation-problem} with the restriction $\vect{x}|_K \equiv 1 \slash |V|^{1/2}$ and $\vect{x}|_M \equiv -1 \slash |V|^{1/2}$ \label{alg-line:earlier-later-influence}\;
		$S, T \leftarrow \emptyset$\;
		\For{$v \in L$}{
			\leIf{$x_v > 0$}{%
				$S{\rm .insert}(v)$
			}
			{%
				$T{\rm .insert}(v)$
			}
		}
		\lIf{$L=V$ \And $|(S \times T) \cap E| < |(T \times S) \cap E| $}{%
			$(S, T) \leftarrow (T, S)$
		}
		$(S,T) \gets {\rm DirectionFix}(G,K,L,M,S,T)$ \label{alg-line:main-top-end-acyc-split} \tcp*{see Algorithm \ref{alg:direction-fix}}
		Replace $L$ with $S \prec T$ in ${\rm VertPrecList}$\;
	}
	\Return $[u \text{ \textbf{for} } u \in U \text{ \textbf{for} } U \in {\rm VertPrecList}]$ \;
\end{algorithm}

We note that the algorithm indeed produces a topological order as it maintains, via the vertex precedence list, a topological order of the coarsened graph (along the partition given by the vertex precedence list) and said coarsened graph is canonically isomorphic to the initial graph when the algorithm terminates. The latter is due to the fact that each set of vertices in the vertex precedence list consists of exactly a single vertex when the algorithm terminates.

The notable difference of Algorithm \ref{alg:spec-top-order-main} to the recursive splitting algorithm of Schamberger--Wierum \cite{schamberger2004locality} lies in Line \ref{alg-line:earlier-later-influence}, where the bi-partition takes into account the previous bi-partitions and their effect on locality.

\begin{remark} \label{rem:recursive-splitting-weight-balance-kway}
	The spectral topological order algorithm, Algorithm \ref{alg:spec-top-order-main}, can be seen as a refinement of a recursive acyclic bi-partitioning algorithm. As such, it (or modifications thereof) can be used to generate more balanced acyclic bi-partitions and even $k$-way acyclic partitions with individually prescribed sizes for each part of the partition.
\end{remark}

%
%
%

\subsection{Evaluation} \label{sec:eval-spec-top-order}

In this section, we will demonstrate that Algorithm \ref{alg:spec-top-order-main} produces topological orders with excellent locality both from a data-centric and a temporal point of view. To this end, we compare our algorithm against ones from the literature on several graph classes using various metrics.

\subsubsection{Algorithms} \label{sec:topOrder-baselines}

We compare our spectral topological order algorithm, Algorithm \ref{alg:spec-top-order-main} with $c = \frac{1}{2|E|}$, and the classic variant ($c=0$), with the following algorithms from the literature:

\begin{itemize}
	\item depth-first search,
	\item breadth-first search with minimum-out-degree ordering,
	\item adapted Cuthill--Mckee \cite{cuthill1969reducing},
	\item adapted Gorder \cite{wei2016speedup} with (recommended) window size of $5$, and
	\item recursive applications of dagP\footnote{Minor modifications were made to recursively bisect using dagP, as it does not allow splitting a graph with less than three nodes
		and would cause a segmentation fault for subgraphs with no edges. Additionally, with loose balance constraints, dagP may assign all nodes to one
		part on small graphs to achieve a zero edge-cut; we therefore enforce a stricter weight imbalance of 0.3.} \cite{herrmann2017acyclic, herrmann2019multilevel} similar\footnote{More precisely, replace lines \ref{alg-line:main-top-begin-acyc-split} through \ref{alg-line:main-top-end-acyc-split} with $(S, T) \gets {\rm dagP}(G|_{M})$.} to Algorithm \ref{alg:spec-top-order-main}.
\end{itemize}

The final algorithm does not exist \emph{per se} in the literature, but may be seen as a na\"ive adaptation of the method by Schamberger--Wierum \cite{schamberger2004locality} to directed acyclic graphs using the state-of-the-art acyclic partitioner dagP.

\subsubsection{Metrics} \label{sec:topOrder-metric}

We are given a finite directed acyclic graph $G=(V,E)$ and a topological order $\preceq$ on $G$, which we shall represent as an enumeration $\sigma$ of the vertices in this section, that is, a bijection $\sigma: V \to \{0,1,\dots , |V| -  1\}$.

There are several metrics to evaluate the `locality' of the topological order both from a data-centric view and from a temporal point of view. Most notably, there is the \emph{bandwidth} \cite{diaz2002survey}, \emph{reuse distance} also known as \emph{LRU-stack distance} \cite{mattson1970evaluation}, and \emph{cut width} \cite{diaz2002survey}. In the literature, this type of problem is known as a graph-layout problem. We refer to the surveys \cite{chung1988labelings, diaz2002survey, petit2013addenda}.

\paragraph{Edge-length distribution}

The length $\omega(e)$ of an edge $e=(u,v) \in E$ is defined as $\sigma(v) - \sigma(u)$. We note that this quantity is (strictly) positive. The edge-length distribution is the collection $\omega(e), e \in E$.

This distribution gives rise to data-locality metrics. For example, we note that the maximum over this distribution is known as the \emph{bandwidth} and minimising the sum of the distribution as the \emph{minimum-linear-arrangement} problem.

\paragraph{Reuse-distance distribution}

In order to define reuse distance, one should think of each vertex producing an output which then gets accessed by its out-neighbours. This produces an access pattern, where we can measure the number of distinct memory accesses between any two accesses to the same piece of memory (output of a vertex). A precise definition of the access pattern and the reuse-distance distribution may be found in Algorithms \ref{alg:access-pattern} and \ref{alg:reuse-distance-distribution}. This distribution gives rise to types of temporal locality.

\begin{algorithm}[!htbp]
	\DontPrintSemicolon
	\SetNlSty{textsc}{}{}
	\SetAlgoNlRelativeSize{-1}
	\caption{Access pattern.\label{alg:access-pattern}}
	\KwData{A directed graph $G=(V,E)$ and a topological order $\preceq$.}
	\KwResult{An access pattern $A$.}
	\BlankLine
	$A \leftarrow \emptyset$\;
	\For{$v\in V$ \emph{in topological order} $\preceq$}{
		\For{$u$ \emph{in-neighbour of} $v$ \emph{in topological order} $\preceq$}{
			$A$.push\_back($u$)\;
		}
		$A$.push\_back($v$)\;
	}
	\Return $A$\;
\end{algorithm}

\begin{algorithm}[!htbp]
	\DontPrintSemicolon
	\SetNlSty{textsc}{}{}
	\SetAlgoNlRelativeSize{-1}
	\caption{Reuse-distance distribution.\label{alg:reuse-distance-distribution}}
	\KwData{An access pattern $A$.}
	\KwResult{Corresponding reuse-distance distribution $D$.}
	\BlankLine
	$D \gets \emptyset$\;
	\For{$i=0, 1, \dots, A$\emph{.size}$()-1$}{
		\If{$\exists j < i$ \emph{such that} $A[j]=A[i]$}{
			$k \gets \max \{ j < i \mid A[j]=A[i] \}$\;
			$d \gets |\{v \in V \mid \exists j  \text{ such that } k < j <i \text{ and } A[j]=v \}|$\;
			$D$.push\_back$(d)$\;
		}
	}
	\Return $D$\;
\end{algorithm}

\paragraph{Edge-cut distribution}

The bisection-cut is the number of cut edges of a bisection of the enumeration $\sigma$ of the topological order $\preceq$. That is
\begin{equation}
	\beta(i) = |\{ (u,v) \in E \mid \sigma(u) \le i < \sigma(v) \}|.
\end{equation}
The edge-cut distribution is the collection of the bisection-cuts $\beta(i)$ for $i=0, 1, \dots, |V|-2$.

This distribution gives rise to data- and temporal-locality metrics. For example, we note that the maximum over this distribution is known as the \emph{cut width} and minimising the sum of the distribution as the \emph{minimum-linear-arrangement} problem.

\begin{remark}
	The sum of all bisection-cuts is equal to the sum of all edge lengths, cf.\@ \cite[Obs.\@ 2, p.\@ 3]{diaz2002survey} and \cite{harper1966optimal}.
\end{remark}

\subsubsection{Data set} \label{sec:topOrder-data-set}

We use the same set of real-world directed graphs as in \S\ref{sec:partition-data-set}, and convert them once more into acyclic instances. Given an input adjacency matrix $A$, we consider its strictly upper ($A_U$) and lower ($A_L$) triangular components. The final acyclic matrix, denoted $A'$, is chosen by identifying the component with the higher edge density, with ties being resolved in favour of $A_U$.


\subsubsection{Results} \label{sec:topOrder-results}

\begin{figure}[!htbp]
    \centering
        \includegraphics[width=0.99\textwidth]{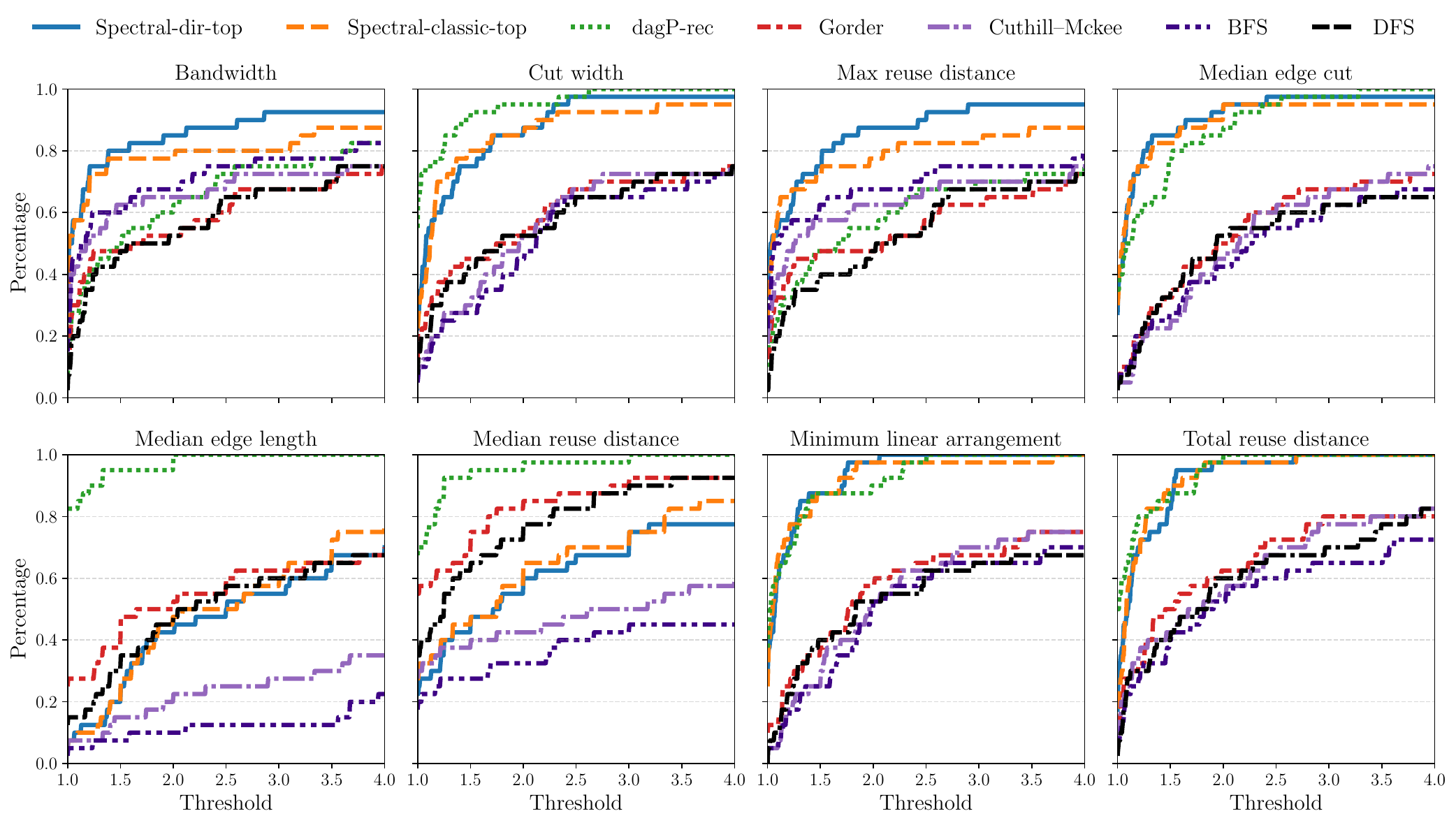}
    \caption{\label{fig:Top_order_perf}  Performance profiles of the algorithms from \S\ref{sec:topOrder-baselines} on the data set from \S\ref{sec:topOrder-data-set} using the various metrics from \S\ref{sec:topOrder-metric}. In each plot, the x-axis represents a threshold and the y-axis the percentage of graphs which are within this threshold times the result of the best algorithm.}
\end{figure}

In Figure \ref{fig:Top_order_perf}, we present performance profiles \cite{dolan2002benchmarking} for the various metrics mentioned in \S\ref{sec:topOrder-metric}.

We observe that the direction-incentivised spectral topological-order algorithm (Spectral-dir-top) outperforms or is on par with its classic variant (Spectral-classic-top) in all metrics. We attribute this to the fact that the direction-incentivised spectral acyclic bi-partition \ref{sec:acyclic-bi-partitioning-algo} is able to preserve more of the original bi-partition, cf.\@ Figure \ref{fig:Partiion_rw_acyc_fix}. Spectral-dir-top further stands out as the single best algorithm for bandwidth, maximum reuse distance, and median edge cut. It is furthermore amongst the best algorithms for minimum linear arrangement and total reuse distance, and near the top in cut width. The recursive application of dagP (dagP-rec) outshines all algorithms on the median edge length, but this comes at the cost of neglecting the tail end of the edge length distribution, e.g., the bandwidth. Similarly, dagP does better on the median reuse distance than any other algorithm, but performs poorly on the maximum reuse distance. Furthermore, dagP is the best on cut width and amongst the best for total reuse distance. Gorder performs well on the median reuse distance. In general, Gorder performs better on temporal-locality metrics for which it was originally designed, but lacks in all other metrics. Cuthill--Mckee and breadth-first search (BFS) perform poorly except for bandwidth and maximum reuse distance where they land in the middle of the pack. Likewise, depth-first search (DFS) performs poorly except for median edge length and median reuse distance where it is above average.

\section{Concluding remarks} \label{sec:conclusion}

We have shown that our direction-incentivised spectral partitioning method, Algorithm \ref{alg:bi-partition}, is able to find bi-partitions with more edges going in one direction. In order to do so, it may increase the edge cut by a small margin. As is well-known, the spectral methods performed better on conductance rather than edge cut and are thus better suited for classification purposes rather than balanced minimum cuts.

We have further demonstrated that spectral partitioning methods (with appropriate fixes) are viable for acyclic partitioning. In particular, when combined with modern multi-level and local search approaches, we believe it to be able to produce high-quality acyclic partitions. This extends to weight-balance constrained $k$-way acyclic partitioning, where we have shown that recursively splitting (for a topological order) leads to a good cut width, cf.\@ Remark~\ref{rem:recursive-splitting-weight-balance-kway}.

Most importantly, we have conclusively shown that our direction-incentivised spectral topological order, Algorithm \ref{alg:spec-top-order-main}, produces exquisite topological orders in terms of data and temporal locality and that the direction incentive improves the locality of the classical analogue even further. In particular, we have found our algorithm to be the best or amongst the best algorithm for bandwidth, maximum reuse distance, median edge cut, minimum linear arrangement, and total reuse distance.

We have further corroborated that the recursively partition method of Schamberger--Wierum \cite{schamberger2004locality} produces good orderings for the bulk of the graph, meaning it performs well on median edge length, edge cut, and reuse distance. We believe the method can be further improved by incorporating ideas mentioned in this paper, specifically by including information of previously cut edges into the subsequent partitioning iterations in order to also address the tail ends of the distributions. We leave this inquiry for future work.

We have thus far avoided talking about time. This is because our implementations are proofs of concepts and little effort has gone into engineering them to be efficient. In practice, spectral methods tend to be a bit slower, but modern implementations are suitably fast \cite{hu2004hsl_mc73, bertrand2013distributed, Naumov16ParSpec, pasadakis23a}. We further point out that Algorithm \ref{alg:spec-top-order-main}, which recursively spectrally splits parts of the graph, generically\footnote{We are assuming here that the spectral bi-partition always finds a $\delta$-balanced bi-partition for a fixed $\delta < 1$, and that its time complexity is upper-bounded by a super-additive and monotone function.} only increases the complexity by a logarithmic factor over the spectral bi-partitioning, Algorithm \ref{alg:bi-partition}.

At last, we bring up the question in what capacity the gains we observed transfer to the various applications, and leave it open as a research topic.



\appendix

\section{Inapproximability: proof of Theorem~\ref{thm:acyclic-bi-partition-inapprox}}
\label{sec:directed-mincut-inapprox}

We first begin with some formal definitions that are required to state and prove Theorem~\ref{thm:acyclic-bi-partition-inapprox}.

\begin{definition}
Given a directed acyclic graph $G=(V,E)$, a disjoint partitioning $V_1 \sqcup V_2 = V$ is called an \emph{acyclic bi-partition} if there is no directed edge $(u,v) \in E$ such that $u \in V_2$ and $v \in V_1$. The \emph{cost} of the bi-partition is the number of edges $(u,v) \in E$ such that $u \in V_1$ and $v \in V_2$.
\end{definition}

\begin{definition} \label{def:balanced-partition}
Given a directed acyclic graph $G=(V,E)$ and a real parameter $0\le \varepsilon < 1$, the acyclic bi-partition $V_1 \sqcup V_2 = V$ is $\varepsilon$-\emph{balanced} if $\max(|V_1|,|V_2|) \le (1+\varepsilon) \cdot \frac{|V|}{2}$. The goal of the \emph{$\eps$-balanced acyclic bi-partition problem} is to find an $\varepsilon$-balanced acyclic bi-partition of minimal cost.
\end{definition}

For the undirected version of the problem, the special case of $\varepsilon=0$ is sometimes also known as the graph bisection problem, and has also been extensively studied, see \cite{feige2002polylogarithmic, racke2008optimal} and references therein/-to.

Given these definitions, we first restate Theorem~\ref{thm:acyclic-bi-partition-inapprox} for completeness.

\thmInapprox*

The theorem provides a particularly interesting contrast to the undirected version of the same problem, which is known to admit an $O(\log{n})$-approximation algorithm~\cite{racke2008optimal, leighton1988approximate}. For the directed case, the only related inapproximability result we are aware of is a simple reduction which only applies when partitioning into a non-constant number of parts~\cite{moreiraAcyclicPartition}.

We split the discussion of the proof into two parts: we first consider the case when the partitions must have equal size ($\varepsilon=0$), and then we show how to extend this to any $0<\varepsilon<1$.

\begin{proof}[Proof for $\varepsilon=0$]
	We provide a reduction from the smallest $k$-edge subgraph problem: given an undirected graph $G_0$ with $N$ vertices and $M$ edges, and an integer $1 \leq k \leq M$, we need to find the smallest subset of vertices that spans at least $k$ edges. It is known that there exists a $\delta'>0$ such that this problem is not approximable in polynomial time to a $n^{1/(\log \log n)^{\delta'}}$ factor, assuming ETH, see \cite{manurangsi2017almost}. Let us select any $\delta > \delta'$.

Given a smallest $k$-edge subgraph problem with $G_0$ and $k$, we develop a directed acyclic graph as follows. We consider two parameters $t=2\cdot M$ and $\ell_0 = t \cdot N+M+2$. The main ingredients of our construction will be \textit{block gadgets}: a set of vertices $v_1,..., v_{\ell}$ such that $(v_i, v_j) \in E$ for all $1 \le i < j \le \ell$. Our construction will ensure that we can trivially find a solution of cost at most $(\ell_0-2)$, and all our block gadgets have size $\ell \ge \ell_0$. This implies that if the vertices $v_1,..., v_{\ell}$ are not all assigned to the same part, then there are at least $(\ell-1) > (\ell_0-2)$ edges cut within the gadget, and hence we already have a higher cost than in the trivial solution. As such, any reasonable solution places each block gadget either entirely into $V_1$ or entirely into $V_2$.

Our directed acyclic graph will consist of the following block gadgets: two block gadgets $B_1$ and $B_2$, a block gadget $B_e$ for each undirected edge $e$ of $G_0$, and a single vertex $v_u$ for each vertex $u$ of $G_0$. We add a single edge from an arbitrary vertex of $B_1$ to an arbitrary vertex of $B_2$. For each vertex $u$ incident to edge $e$ in $G_0$, we add an edge from $v_u$ to an arbitrary vertex of $B_e$. Finally, for all vertices $v_u$ that represent a vertex in $G_0$, we add $t$ distinct edges from $v_u$ to $t$ arbitrary vertices of $B_2$. As for the size of the block gadgets, the number of vertices will be
\begin{itemize}
\item $(M-k) \cdot \ell_0+N+1$ in $B_1$,
\item $k \cdot \ell_0+N+1$ in $B_2$,
\item $\ell_0$ in each $B_e$.
\end{itemize}
Finally, we add $N$ further isolated vertices to the directed acyclic graph. Altogether, the size of the construction is $n=2M \cdot \ell_0 + 4 \cdot N + 2=O(N \cdot M^2)=O(N^5)$.

We have $|B_1|+|B_2|>\frac{n}{2}$, so the two blocks cannot be in the same part. Due to the edge between them, this implies that $B_1$ is in $V_1$ and $B_2$ is in $V_2$. Since $\ell_0 > 2 \cdot N$, in order to have exactly $\frac{n}{2}$ vertices in both parts without splitting a block gadget, we need to assign exactly $k$ block gadgets $B_e$ to $V_1$, and the remaining $(M-k)$ block gadgets $B_e$ to $V_2$. We can then assign any desired subset of the $N$ vertices $v_u$ to $V_1$, and the rest to $V_2$, since we can balance it afterwards with the $N$ isolated vertices.

Selecting the $k$ gadgets $B_e$ to assign to $V_1$ corresponds to selecting $k$ edges in $G_0$. If an edge $e$ in $G_0$ is selected this way, and it is incident to a vertex $u$ in $G_0$, then this implies that $v_u$ must also be in $V_1$, since $v_u$ has an edge to a vertex in $B_e$. As such, the vertices $v_u$ corresponding to all the endpoints of the $k$ selected edges must be in $V_1$. All these vertices $v_u$ will have $t$ outgoing edges to $B_2$ in $V_2$, and some further edges to the $B_e$ incident to $u$; some of these might also be in $V_2$. On the other hand, if a vertex $u$ has all of its incident edges in $V_2$, then no reasonable solution will place $v_u$ in $V_1$, since this results in a lot of cut edges; indeed, any such solution can be easily improved by exchanging $v_u$ with an isolated vertex.

This means that if a set of $k$ selected edges in $u_0$ spans $c$ vertices, then the resulting bi-partition will have cost $c \cdot t$ plus the sum of the degree of the spanned vertices, i.e.\ cost between $c \cdot t$ and $c \cdot t + M$. With $t=2M$, this cost is always dominated by the term $c \cdot t$. The observation also holds the other way around: if we have a reasonable solution in the directed acyclic graph for the acyclic bi-partition problem (i.e., no blocks are cut, no $v_u$ is placed in $V_1$ unless necessary), and the solution has cost between $c \cdot t$ and $(c+1) \cdot t$ for some $c \in \mathbb{Z}_{\ge 0}$, then this allows us to select a subset of $c$ vertices in $G_0$ that induce at least $k$ edges.

More formally, let ${\rm OPT_{SkES}}$ and ${\rm OPT_{ABP}}$ denote the optimum costs in the original smallest $k$-edge subgraph problem and the derived acyclic bi-partition problem, respectively. As discussed before, we have ${\rm OPT_{ABP}} \le t \cdot {\rm OPT_{SkES}} + M$. Assume now that we have a polynomial-time algorithm that approximates the acyclic bi-partition problem to a $n^{1/(\log \log n)^{\delta}}$ factor. We can assume that this algorithm returns a reasonable solution (no blocks are cut, no $v_u$ is placed in $V_1$ unless necessary), otherwise we begin by replacing it by a reasonable solution as discussed above; this only further decreases its cost. Assume that the algorithm returns a solution of cost ${\rm SOL_{ABP}}$. Consider the derived solution of smallest $k$-edge subgraph problem, where we select the $k$ edges whose edge gadgets are in $V_1$; let us denote its cost by ${\rm SOL_{SkES}}$. Recall that by assumption, we have
\begin{equation}
{\rm SOL_{ABP}} \le n^{1/(\log \log n)^{\delta}} \cdot {\rm OPT_{ABP}} \, .
\end{equation}
Recall from before that ${\rm SOL_{ABP}} \ge t \cdot {\rm SOL_{SkES}}$, and hence
\begin{equation}
t \cdot {\rm SOL_{SkES}} \le n^{1/(\log \log n)^{\delta}} \cdot (t \cdot {\rm OPT_{SkES}} + M) \, .
\end{equation}
With $t>M$, we also get
\begin{equation}
{\rm SOL_{SkES}} \le n^{1/(\log \log n)^{\delta}} \cdot ({\rm OPT_{SkES}} + 1) \le 2 \cdot n^{1/(\log \log n)^{\delta}} \cdot {\rm OPT_{SkES}} \, ,
\end{equation}
where the second inequality follows easily for ${\rm OPT_{SkES}} \ge 1$. Recall that there is a constant $c_0$ such that $n \le c_0 \cdot N^5$, so we can upper-bound $n^{1/(\log \log n)^{\delta}}$ by $(c_0 \cdot N^5)^{1/(\log \log N)^{\delta}}$. Then, $(2c_0 \cdot N^5)^{1/(\log \log N)^{\delta}}$ is in turn upper-bounded by $N^{6/(\log \log N)^{\delta}}$ for sufficiently large $N$. This implies
\begin{equation}
{\rm SOL_{SkES}} \le N^{6/(\log \log N)^{\delta}} \cdot {\rm OPT_{SkES}} \, .
\end{equation}
With $\delta > \delta'$ and $N$ large enough, $N^{6/(\log \log N)^{\delta}}$ is upper-bounded by $N^{1/(\log \log N)^{\delta'}}$, hence
\begin{equation}
{\rm SOL_{SkES}} \le N^{1/(\log \log N)^{\delta'}} \cdot {\rm OPT_{SkES}} \, .
\end{equation}
As such, our algorithm allows to approximate the smallest $k$-edge subgraph problem in polynomial time to a factor that is not possible if ETH holds.
\end{proof}

The same reduction can easily be adapted to the case of more flexible balance constraints.

\begin{proof}[Proof for $0<\varepsilon<1$]
In order to adapt the proof above to the $\varepsilon$-balanced case, we only need to adjust the size of $B_1$ and $B_2$ to ensure that we are still forced to place $k$ edge gadgets into $V_1$, i.e., we can only fit $B_2$, at most $(M-k)$ edge gadgets and $N$ further vertices into $V_2$. For simplicity, let $s$ denote the total size of the edge and vertex gadgets and isolated vertices, i.e., $s=M \cdot \ell_0+2N$. Let $s_0$ denote the size of $(M-k)$ edge gadgets plus $N$ further vertices, i.e., $s_0=(M-k) \cdot \ell_0 + N$. We need to ensure that
\begin{equation}
|B_2|+ s_0 \le (1+\varepsilon) \cdot \frac{n}{2} \le |B_2|+ s_0 + N \, ,
\end{equation}
while $|B_1|+|B_2| > (1+\varepsilon) \cdot \frac{n}{2}$ also remains true. For this, we can first select $n$ large enough such that $s < (1-\varepsilon) \cdot \frac{n}{2}$. We can then set $|B_2|= \lfloor (1+\varepsilon) \cdot \frac{n}{2} \rfloor - s_0$, and finally $|B_1|=n-|B_2|-s$.

Note that these choices also implicitly ensure for $V_1$ that $|B_1| +k \cdot \ell_0 + N \le (1+\varepsilon) \cdot \frac{n}{2}$: after substituting our choices of $|B_1|$, $|B_2|$, $s_0$ and $s$, we get $n \le (1+\varepsilon) \cdot \frac{n}{2} + \lfloor (1+\varepsilon) \cdot \frac{n}{2} \rfloor$, which holds for $n$ large enough.
\end{proof}

\begin{remark}
	If one assumes the stronger Gap-ETH conjecture instead of ETH, then the same reduction shows an inapproximability to an $n^{o(1)}$ factor~\cite{manurangsi2017almost}.
\end{remark}

\section{List of graphs}
\label{sec:graph-tables}

\begin{longtblr}[
theme = ams-theme,
caption = {All graphs from the SuiteSparse Matrix Collection~\cite{Davis11} and generated by the HyperDAG\_DB~\cite{HyperDAG-DB, papp24a} used in the paper with source domain, number of vertices, number of edges, and percentage of symmetric edges listed.},
label = {table:all-graphs},
]{ colspec = { Q[l,m] | Q[l,m] | Q[r,m] | Q[r,m] | Q[r,m] }, row{1} = {font=\bfseries}, rowhead = 1, rowfoot = 0}
Graph & Domain & Vertices & Edges & Sym.\@ Edges
\\ \hline \hline arc130 & materials science & 130 & 1,282 & 75.9\%		
\\ \hline b2\_ss & chemical process & 1,089 & 4,228	& 1.5\%				
\\ \hline barth & structural simulation & 6,691 & 26,439 & 0.0\%		
\\ \hline barth4 & structural simulation & 6,019 & 23,492 & 0.0\%		
\\ \hline barth5 & structural simulation & 15,606 & 61,484 & 0.0\%		
\\ \hline bayer03 & chemical process & 6,747 & 56,196 & 0.3\%			
\\ \hline bp\_1600 & optimisation & 822 & 4,841 & 1.1\%                 
\\ \hline CG\_bcsstk05 & scheduling & 11,041 & 23,116 & 0.0\%           
\\ \hline CG\_nos1 & scheduling & 10,256 & 19,384 & 0.0\%     			
\\ \hline circuit204 & circuit simulation & 1,020 & 5,883 & 43.8\%		
\\ \hline conf5\_0-4x4-10 & optimisation & 3,072 & 119,808 & 92.3\%     
\\ \hline cz5108 & fluid dynamics & 5,108 & 51,412 & 43.4\%				
\\ \hline dw2048 & electromagnetics & 2,048 & 10,114 & 98.5\%			
\\ \hline dw4096 & electromagnetics & 8,192 & 41,746 & 96.3\%   		
\\ \hline epb0 & thermodynamics & 1,794 & 7,764 & 50.1\%				
\\ \hline foldoc & networks & 13,356 & 120,238 & 47.9\%                 
\\ \hline fp & electromagnetics & 7,548 & 834,222 & 76.8\%              
\\ \hline fpga\_dcop\_15 & circuit simulation & 1,220 & 5,892 & 81.8\%	
\\ \hline fs\_541\_1 & 3D problem & 541 & 4,285	& 68.3\%				
\\ \hline g7jac050sc & economic problem & 14,760 & 145,157 & 3.2\%      
\\ \hline gemat12 & power network & 4,929 & 33,111 & 0.1\%				
\\ \hline gre\_512 & chip simulation & 512 & 2,192 & 0.0\%				
\\ \hline lhr14c & chemical process & 14,270 & 307,858 & 0.7\%			
\\ \hline lshp1270 & thermodynamics & 1,270 & 8,668 & 100.0\%			
\\ \hline mahindas & economic problem & 1,258 & 7,682 & 1.7\%			
\\ \hline mark3jac020sc & economic problem & 9,129 & 52,883 & 6.1\%     
\\ \hline olm5000 & fluid dynamics & 5,000 & 19,996 & 66.7\%            
\\ \hline orani678 & optimisation & 2,529 & 90,158 & 7.1\%              
\\ \hline p2p-Gnutella06 & networks & 8,717 & 31,525 & 0.0\%			
\\ \hline pores\_2 & fluid dynamics & 1,224 & 9,613 & 61.2\%            
\\ \hline POW3SPMV\_bcspwr05 & scheduling & 5,904 & 9,297 & 0.0\%       
\\ \hline psmigr\_3 & economic problem & 3,140 & 543,160 & 47.9\%       
\\ \hline qh882 & networks & 882 & 3,354 & 92.6\%                       
\\ \hline rajat06 & circuit simulation & 10,922 & 28,922 & 0.0\%		
\\ \hline rdist3a & chemical process & 2,398 & 61,896 & 14.0\%	  		
\\ \hline rw5151 & optimisation & 5,151 & 20,199 & 49.0\%               
\\ \hline SPMV\_west0497 & scheduling & 4,448 & 5,181 & 0.0\%           
\\ \hline ted\_A & thermodynamics & 10,605 & 424,587 & 56.6\%           
\\ \hline utm5940 & electromagnetics & 5,940 & 83,842 & 52.9\%			
\\ \hline watt\_1 & fluid dynamics & 1,856 & 11,360	& 98.7\%			
\end{longtblr}

\begin{landscape}
\pagestyle{empty}
\section{Individual results for topological orders}
\label{sec:full_top_order_res}

    \vspace*{\fill}
\begin{figure}
    \includegraphics[width=1\linewidth]{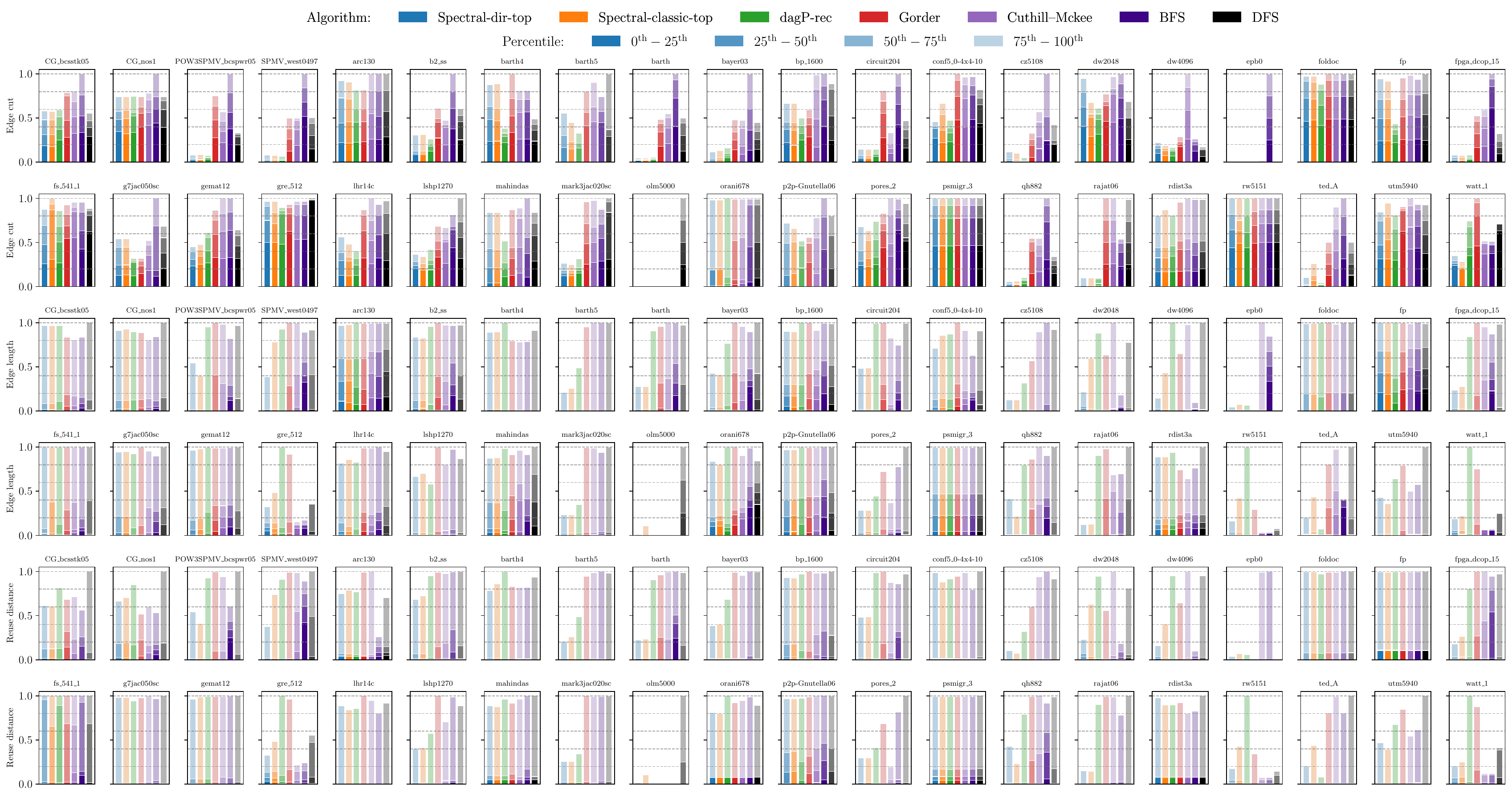}
    \caption{Edge-length, reuse-distance, and edge-cut distributions of the topological orders generated by the algorithms in \S\ref{sec:topOrder-baselines} on each individual graph from the data set in \S\ref{sec:topOrder-data-set}. For each graph and metric, the values have been rescaled such that the worst performing algorithm takes the value $1$.}
    \label{fig:TopOrder-all-graphs}
\end{figure}
    \vspace*{\fill}
\end{landscape}

\bibliography{refs}{} \bibliographystyle{alpha}
\end{document}